\RequirePackage{filecontents}

\RequirePackage{snapshot}
\documentclass[aps,pra,reprint,showpacs,secnumarabic,superscriptaddress]{preprint}

\let\Paragraph=\paragraph
\AtBeginDocument{\renewcommand{\paragraph}[1]{\Paragraph{#1}}}
\renewcommand\documentclass[2][]{}
\documentclass[aps,pra,reprint,twocolumn,showpacs,floatfix,superscriptaddress]{revtex4-1}

\usepackage[american]{babel}
\usepackage{amssymb}
\usepackage{amsmath}
\usepackage[T1]{fontenc}
\usepackage{txfonts}
\usepackage{textcomp}
\usepackage{dsfont}
\usepackage{braket}
\usepackage{isomath}
\usepackage{microtype}
\makeatletter
\renewcommand*{\vec}[1]{\ifcat a#1\mathbfit{#1}\else\boldsymbol{#1}\fi}
\makeatother
\usepackage{graphicx}
\graphicspath{{figures/}}
\usepackage{color}
\usepackage{hyperref}
\hypersetup{breaklinks=true,%
  pdfauthor={Sven Ahrens, Heiko Bauke, Christoph H. Keitel, and Carsten M{\"u}ller},%
  pdftitle={Kapitza-Dirac effect in the relativistic regime}}
\newcommand*{\I}{\mathrm{i}}
\newcommand*{\e}{\mathrm{e}}
\newcommand*{\D}{\mathrm{d}}
\newcommand*{\E}{\mathcal{E}}
\newcommand*{\trans}[1]{#1^{\mathsf{T}}}
\DeclareMathOperator{\sign}{sign}
\DeclareMathOperator{\tr}{tr}

\begin{document}

\title{Kapitza-Dirac effect in the relativistic regime}

\author{Sven Ahrens}
\author{Heiko Bauke}
 \email{heiko.bauke@mpi-hd.mpg.de}
\author{Christoph H. Keitel}
\affiliation{Max-Planck-Institut f{\"u}r Kernphysik,
  Saupfercheckweg~1, 69117~Heidelberg, Germany}
\author{Carsten M{\"u}ller}
 \email{mueller@tp1.uni-duesseldorf.de}
\affiliation{Max-Planck-Institut f{\"u}r Kernphysik,
   Saupfercheckweg~1, 69117~Heidelberg, Germany}
\affiliation{Institut f{\"u}r Theoretische Physik~I,
   Heinrich-Heine-Universit{\"a}t D{\"u}sseldorf, Universit{\"a}tsstra{\ss}e~1,
   40225~D{\"u}sseldorf, Germany}

\date{\today}

\begin{abstract}
  A relativistic description of the Kapitza-Dirac effect in the
  so-called Bragg regime with two and three interacting photons is
  presented by investigating both numerical and perturbative solutions
  of the Dirac equation in momentum space. We demonstrate that
  spin-flips can be observed in the two-photon and the three-photon
  Kapitza-Dirac effect for certain parameters. During the interaction
  with the laser field the electron's spin is rotated, and we give
  explicit expressions for the rotation axis and the rotation angle.
  The off-resonant Kapitza-Dirac effect, that is, when the Bragg
  condition is not exactly fulfilled, is described by a generalized Rabi
  theory.  We also analyze the in-field quantum dynamics as obtained
  from the numerical solution of the Dirac equation.
\end{abstract}


\maketitle

\section{Introduction}
\label{sec:intro}

The diffraction of electrons at a standing wave of light is referred to
as the Kapitza-Dirac effect
\cite{Kapitza:Dirac:1933:effect,Batelaan:2007:Illuminating_Kapitza-Dirac}.
It is the counterpart process of the usual diffraction of light at a
material grating.  The observation of the Kapitza-Dirac effect seemed to
be feasible with the advent of the laser
\cite{Schwarz:etal:1965:direct_observation,Schwarz:etal:1968:second_observation_claim}
but its experimental realization was refuted shortly after the claim of
detection
\cite{Pfeiffer:1968:Experimentelle_Prufung,Takeda:Matsui:1968:Electron_Reflection,Bartell:1968:comment_on_schwarz}.
The first observation of the Kapitza-Dirac effect in the so-called Bragg
regime was achieved by employing atoms
\cite{Adams:etal:1994:Atom_optics,Freyberger:etal:1999:Atom_Optics} in
1986 \cite{Gould:etal:1986:Diffraction_of_atoms_by_light} and two years
later in the so-called diffraction regime
\cite{Martin:etal:1988:Bragg_scattering}.  The realization of the
Kapitza-Dirac effect with electrons was achieved in 1988
\cite{Bucksbaum:etal:1988:Kapitza-Dirac_Effect} in the diffraction
regime.  The scattering of electrons in the Bragg regime was
demonstrated in a precise and sophisticated experimental setup in 2001
\cite{Freimund:etal:2001:Observation_Kapitza-Dirac,Freimund:Batelaan:2002:Bragg}.
Among all experiments this comes closest to the diffraction process as
proposed originally by Kapitza and Dirac
\cite{Kapitza:Dirac:1933:effect}.

The latter experiment was performed at nonrelativistic laser intensities
with a laser field parameter \cite{DiPiazza:etal:2012:review}
$\xi_0=eE_0/(mc\omega)=5\times10^{-6}$ (with $-e$ denoting the
electron's charge and $m$ its mass, laser electric field amplitude $E_0$
and angular frequency $\omega$, and the speed of light $c$), with
nonrelativistic electron momenta $|\vec p| = 0.04\,m c$, and at
nonrelativistic photon energies $\E=5\times10^{-6}\,mc^2$, raising the
question \cite{Freimund:etal:2001:Observation_Kapitza-Dirac} of how the
Kapitza-Dirac effect might be modified in the relativistic regime.
Higher intensities
\cite{Yanovsky:etal:2008:Ultra-high_intensity,ELI,Mourou:Tajima:2011:More_intense}
and shorter wavelengths
\cite{Altarell:etal:2007:XFEL,Emma:etal:2010:First_lasing,McNeil:Thompson:2010:XFEL}
are indeed available nowadays, demanding a relativistic theory of
Kapitza-Dirac scattering within the framework of the Dirac theory which
also accounts for the electron's spin degree of freedom.  Beside
considerations of the Kapitza-Dirac effect with adiabatic pulse turn-on
\cite{Fedorov:1974:Stimulated_scattering}, two electrons
\cite{Sancho:2010:two-particle_Kapitza-Dirac,Sancho:2011:Bragg_two-particle_Kapitza-Dirac},
and perturbative solutions
\cite{Gush:1971:Electron_Scattering,Efremov:Fedorov:1999:Classical_and_quantum,Efremov:Fedorov:2000:Wavepacket_theory,Smirnova:etal:2004:two_color_Kapitza-Dirac},
relativistic investigations have been described in
\cite{Haroutunian:Avetissian:1975:analogue_KGE,Federov:McIver:1980:compton_scattering}
based on the Klein-Gordon equation and therefore neglecting the spin.
Freimund and Batelaan raised the question of whether the electron spin
is affected in the Kapitza-Dirac effect, but found a vanishingly small
spin-flip probability in the investigated parameter regime by simulating
the Bargmann-Michel-Telegdi equations
\cite{Freimund:Batelaan:2003:Stern-Gerlach}. Another derivation which
was carried out by Rosenberg \cite{Rosenberg:2004:Extended_theory}
solved the Pauli equation perturbatively in a second-quantized field in
the diffraction regime and also found only tiny spin effects.  The
impact of the electron's spin has also been investigated with respect to
free-electron motion
\cite{Krekora:Su:Grobe:2001:spin_of_relativistic_electrons,*Walser:Urbach:Hatsagortsyan:Hu:Keitel:2002:spin_and_radiation_in_intense_laser_fields},
bound-electron dynamics \cite{Hu:Keitel:1999:Spin_Signatures}, atomic
photoionization
\cite{Faisal:Bhattacharyya:2004:helium_double_ionization_spin_assymetry},
and Compton and Mott
\cite{Szymanowski:etal:1998:laser_scattering,*Ivanov:etal:2004:polarization,*Ehlotzky:etal:2009:Fundamental_processes}
scattering in strong plane-wave laser fields.  In addition collapse and
revival spin dynamics has been put forward in strongly laser-driven
electrons \cite{skoromnik:etal:2013:collapse_revival} and notable spin
signatures have been found in laser-induced ionization
\cite{klaiber:etal:2013:spin_in_ionization}.  Our recent publication on
the Kapitza-Dirac effect \cite{Ahrens:etal:2012:spin_kde} solves the
Dirac equation numerically and perturbatively in the Bragg regime and
demonstrates pronounced spin dynamics in the Kapitza-Dirac effect
involving three photons.  In the present paper, we elaborate our theory
in detail and discuss the possibility of spin dynamics for Kapitza-Dirac
scattering with only two interacting photons.

This article is organized as follows.  In
Sec.~\ref{sec:energy-momentum-conservation}, we specify the Bragg and
the diffraction regimes and characterize the interaction in the Bragg
regime employing a classical picture. The quantum equations of motions
(namely, the Pauli and Dirac equations) are considered in
Sec.~\ref{sec:waveequations}, where it is shown that for the setup of
the Kapitza-Dirac effect they take a particular simple form in momentum
space.  Based on these equations, we study the Kapitza-Dirac effect with
two interacting photons in Sec.~\ref{sec:two_photon_KDE} numerically and
analytically and demonstrate the occurrence of spin flips at
relativistic electron momenta. We show that the diffraction probability
is independent of the spin orientation of the incident electron beam.
The spin orientation of the diffracted electrons is rotated as a result
of the interaction with the laser field. We also discuss the resonance
peak structure of the diffraction within a generalized Rabi theory. In
Sec.~\ref{sec:three-photon-KDE}, we investigate the Kapitza-Dirac effect
with three interacting photons in close analogy to the two-photon
case. We analyze the in-field quantum dynamics of the Kapitza-Dirac
effect and we discuss the tilt of the axis about which the electron spin
is rotated.

\section{Semi-classical considerations}
\label{sec:energy-momentum-conservation}

Depending on the parameters of the laser and the incident electron, the
scattering dynamics of the Kapitza-Dirac effect happens in different
regimes.  Phenomenologically one refers to the Bragg regime if the
electron is scattered into a single diffraction order and one refers to
the diffraction regime if many diffraction orders can be reached.
Employing an argument based on the time-energy uncertainty relation
\cite{Batelaan:2000:Kapitza-Dirac,Freimund:Batelaan:2002:Bragg} one can
show that if the interaction time of the electron with the laser is
short, the electron is diffracted into several diffraction orders.  Long
interaction times, however, permit dynamics in the Bragg regime
\cite{Gush:1971:Electron_Scattering}.  In the Bragg regime the electron
can be diffracted only if it fulfills the classical energy and
momentum conservation.  This article focuses on the Kapitza-Dirac effect
in the Bragg regime.

\begin{figure}
  \centering
  \includegraphics{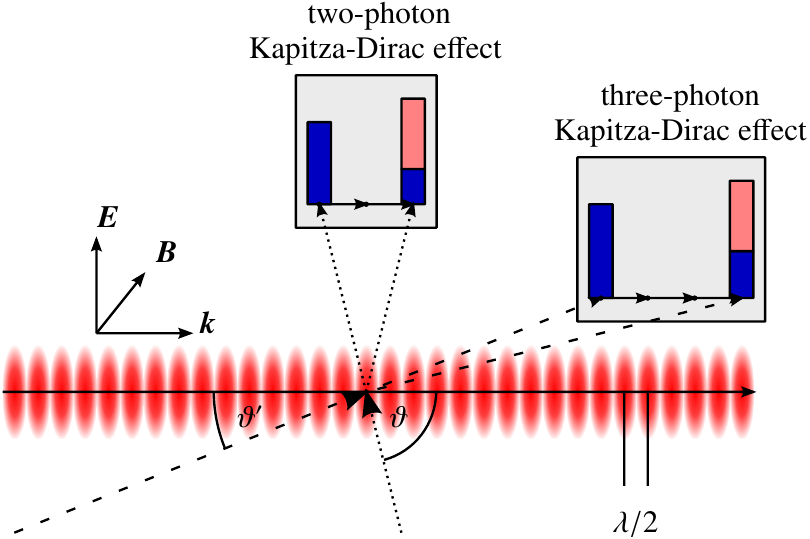}
  \caption{(Color online) Setup of the two-photon and the three-photon
    Kapitza-Dirac effects. If the electron is diffracted, it may pick up
    two photon momenta or three photon momenta, respectively, from the
    laser field.  Directions of the electron beams are indicated by
    dotted (two-photon Kapitza-Dirac effect) and dashed (three-photon
    Kapitza-Dirac effect) lines.  The gray panels emblematize screens,
    which display the intensity of the scattered and unscattered
    electron beams (colored bars), depending on the electron momentum in
    the laser propagation direction. In both Kapitza-Dirac effects spin
    flips may occur, indicated by the light red bar of the diffraction
    pattern as compared to the dark blue bars of the unflipped electron
    beam intensity. According to the considerations of
    Secs.~\ref{sec:energy-momentum-conservation}
    and~\ref{sec:two_photon_KDE}, the electron enters the laser almost
    perpendicular to the laser's propagation direction for the
    two-photon Kapitza-Dirac effect with spin-flip dynamics, implying
    $\vartheta \approx 90$\textdegree. For the three-photon
    Kapitza-Dirac effect, however, the energy-momentum conservation
    constraint \eqref{eq:p_k_relativistic} results in relativistic
    momenta of the electron in the laser propagation direction such that
    typically $\vartheta' \ll90$\textdegree.}
  \label{fig:experimental-setup}
\end{figure}

The Kapitza-Dirac effect can be viewed as a scattering process where the
electron either absorbs or emits photons from or into the
counter-propagating laser beams.  We denote the number of absorbed
photons from the left- (right-) traveling laser beam by $n_l$
($n_r$). Negative values of $n_l$ or $n_r$ correspond to photon emission
of the electron into the left or the right laser beam.  Following this
notion, classical momentum conservation requires
\begin{equation}
  \label{eq:momentum-conservation}
  \vec p_{\mathrm{out}} = \vec p_{\mathrm{in}} + n_r \vec k - n_l \vec k\,.
\end{equation}
Note that Gaussian units are used in this article and we set $\hbar=1$.
Furthermore, energy conservation implies
\begin{equation}
  \label{eq:energy-conservation}
  \E(\vec p_{\mathrm{out}}) = \E(\vec p_{\mathrm{in}}) + n_r c k + n_l c k \,,
\end{equation}
with $\vec p_{\mathrm{in}}$ and $\vec p_{\mathrm{out}}$ denoting the
electron's momentum before and after interaction with the laser, the
electron's kinetic energy $\E(\vec p)$, and the photon momentum $\vec
k$. For convenience, we separate the electron's momentum into components
in the laser propagation direction $p_k$, the laser polarization
direction $p_E$, and the direction of the laser magnetic field $p_B$ (see
Fig.~\ref{fig:experimental-setup}).
Applying the nonrelativistic energy-momentum relation
\begin{equation}
  \E^\mathrm{nr}(\vec p) = \frac{\vec p^2}{2 m} \,,
\end{equation}
the solution of Eq.~\eqref{eq:energy-conservation} with
respect to $p_{k,\mathrm{in}}$ yields
\begin{equation}
  \label{eq:p_k_nonrelativistic}
  p_{k,\mathrm{in}}^\mathrm{nr} = 
  -\frac{n_r - n_l}{2} k + \frac{n_r + n_l}{n_r - n_l} m c\,,
\end{equation}
whereas Eq.~\eqref{eq:energy-conservation} with the relativistic
energy-momentum relation
\begin{equation}
  \label{eq:relativistic_energy_momentum_relation}
  \E^\mathrm{r}(\vec p) = \sqrt{m^2 c^4 + c^2 \vec p^2}
\end{equation}
yields
\begin{multline}
  \label{eq:p_k_relativistic}
  p_{k,\mathrm{in}}^\mathrm{r} = 
  - \frac{n_r - n_l}{2} k \\ + 
  \frac{n_r - n_l}{|n_r - n_l|} \frac{n_r + n_l}{2} 
  \sqrt{k^2 - 
    \frac{m^2 c^2 + ({p_{B,\mathrm{in}}^\mathrm{r}})^2 + 
      ({p_{E,\mathrm{in}}^\mathrm{r}})^2}{n_r n_l}}\,.
\end{multline}
Note that the solution \eqref{eq:p_k_relativistic} is real-valued and
finite only if $n_r$ and $n_l$ are non-zero and have opposite
signs. This means that the Kapitza-Dirac effect in the Bragg regime
requires that at least one photon is absorbed from one laser beam and at
least one photon is emitted into the counterpropagating laser beam.  For
$n_r + n_l = 0$ the electron's energy \eqref{eq:energy-conservation} is
conserved and, therefore, the scattering is elastic; otherwise it is
inelastic.

\begin{figure}
  \centering
  \includegraphics{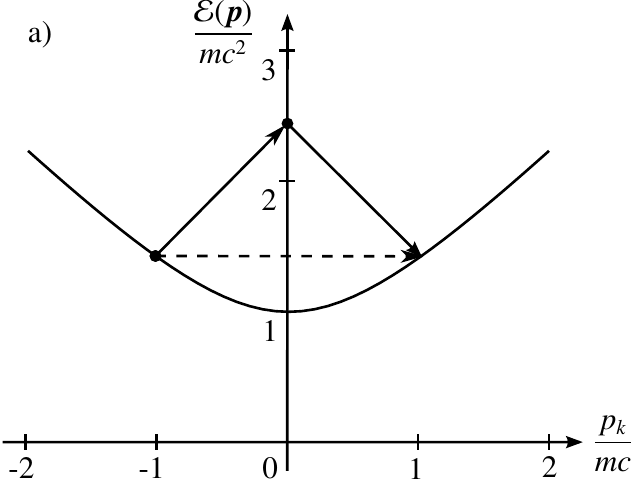}\\[3ex]
  \includegraphics{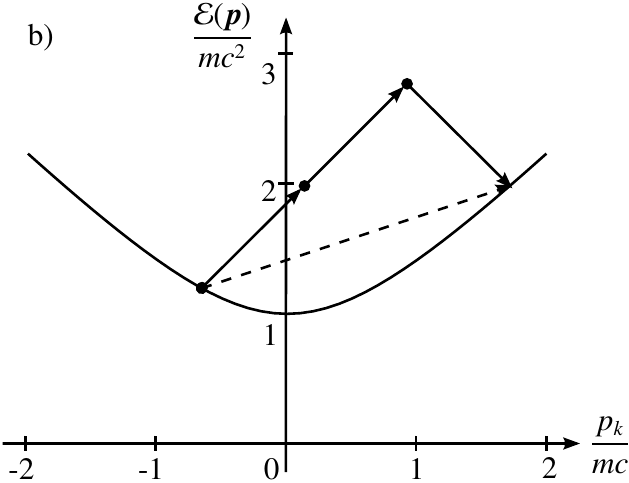}
  \caption{Sketch of the energy and momentum conservation in the
    Kapitza-Dirac effect.  The electron must reside on its relativistic
    energy-momentum relation
    \eqref{eq:relativistic_energy_momentum_relation} (solid hyperbola)
    before as well as after the interaction, and the laser can transfer
    only energy and momentum quanta which fulfill $\Delta\mathcal{E}=c
    k$ (diagonal arrows).  In the two-photon Kapitza-Dirac effect (a),
    one photon is absorbed and one photon is emitted, whereas in the
    three-photon Kapitza-Dirac effect (b) two photons are absorbed and
    one photon is emitted.  In the latter process the energy of the
    electron changes during diffraction as indicated by the
    nonhorizontal dashed line.}
  \label{fig:KDE_energy_momentum_conservation}
\end{figure}

Equation~\eqref{eq:p_k_nonrelativistic} as well as
Eq.~\eqref{eq:p_k_relativistic} yield
$p_{k,\mathrm{out}}^\mathrm{nr/r}=-p_{k,\mathrm{in}}^\mathrm{nr/r}$ for
elastic electron scattering corresponding to ``The reflection of
electrons from standing light waves'' proposed by Kapitza and Dirac
\cite{Kapitza:Dirac:1933:effect}.  Energy-momentum conservation can be
illustrated in an energy-momentum diagram as shown in
Fig.~\ref{fig:KDE_energy_momentum_conservation}(a) for the elastic
two-photon Kapitza-Dirac effect.

In the case of the three-photon Kapitza-Dirac effect, two laser photons
are absorbed ($n_r=2$) and one is emitted ($n_l=-1$). The
energy-momentum diagram of this process is sketched in
Fig.~\ref{fig:KDE_energy_momentum_conservation}(b). Processes with
$n_r=-1$ and $n_l=2$, $n_r=-2$ and $n_l=1$, or $n_r=1$ and $n_l=-2$ are
also possible but not fundamentally different from the Kapitza-Dirac
effect with $n_r=2$ and $n_l=-1$ and therefore not considered here. For
such an inelastic process, energy is transferred from the laser field to
the electron.  This energy transfer is indicated by a dashed line in
Fig.~\ref{fig:KDE_energy_momentum_conservation}(b).  Approaching the
limit $\vec k\to 0$, the dashed line in
Fig.~\ref{fig:KDE_energy_momentum_conservation}(b) is shifted downwards
while preserving its slope until it becomes a tangent of the energy
hyperbola \eqref{eq:relativistic_energy_momentum_relation}.
Consequently, energy-momentum conservation requires a non-vanishing
initial momentum of the electron in the laser propagation direction even
for small photon momenta.  Furthermore, the energy conservation
\eqref{eq:energy-conservation} enforces that at least one of the momenta
$k$, $p_k$, $p_E$, or $p_B$ is relativistic, i.\,e., of the order of or
larger than $mc$.  Consequently, no nonrelativistic limit of
\eqref{eq:p_k_relativistic} exists in this case. For example, the
momenta \eqref{eq:p_k_nonrelativistic} and \eqref{eq:p_k_relativistic}
do not converge even for small laser photon momenta $k$.  Rather we find
with $n_r=2$ and $n_l=-1$ the differing limiting values
\begin{subequations}
  \label{eq:small_k_pk_limits}
  \begin{equation}
    \lim_{k \to  0} p_{k,\mathrm{in}}^\mathrm{nr} = \frac{m c}{3} 
  \end{equation}
  and
  \begin{equation}
    \lim_{k \to  0} p_{k,\mathrm{in}}^\mathrm{r} =
    \sqrt{\frac{m^2c^2+p_E^2 +p_B^2}{8}}\,.
  \end{equation}
\end{subequations}

For $p_{k,\mathrm{in}}^\mathrm{r}$ to be small compared to $mc$ the
number of interacting photons has to go to infinity such that $|n_r -
n_l| \gg 1$ and $|n_r + n_l| \approx 1$.  In this case the slope of the
energy-momentum transfer (dashed lines in
Fig.~\ref{fig:KDE_energy_momentum_conservation}) goes to zero.
Therefore, the touching point of the corresponding tangent that results
in the limit $\vec k \to 0$ lies at small electron momenta in laser
propagation direction for Kapitza-Dirac scattering with a high number of
interacting photons.

\section{Quantum dynamics in momentum space}
\label{sec:waveequations}

In order to study spin effects and the quantum dynamics of the
two-photon and three-photon Kapitza-Dirac effects in a relativistic
setting we will solve the time-dependent Dirac equation.  For two-photon
interactions we will also utilize the nonrelativistic Pauli equation
given that relativistic parameters are not mandatory in this case.  As
we will show in this section, the Pauli equation and the Dirac equation
for the setup of the Kapitza-Dirac effect can be reduced to a system of
ordinary differential equations by transforming them into momentum space

\subsection{Laser setup}
\label{sec:laser}

For the Kapitza-Dirac effect we consider two counterpropagating linearly
polarized lasers of equal intensity and angular frequency $\omega$.  The
vector potential of this laser setup is given by
\begin{equation}
  \label{eq:vector_potential}
  \vec A(\vec x, t) = \vec {\hat A}(t) \cos(\vec k \cdot \vec x) \sin(\omega t) \,,
\end{equation}
where we have introduced the temporal envelope function
\begin{equation}
  \label{eq:turn_on_and_off}
  \vec {\hat A}(t) = \vec {\hat A}_\mathrm{max}\times
  \begin{cases}
    \sin^2 \frac{\pi t}{2\Delta T} & \text{if $0\le t\le \Delta T$,} \\
    1 & \text{if $\Delta T\le t\le T-\Delta T$,} \\
    \sin^2 \frac{\pi (T-t)}{2\Delta T} & \text{if $T-\Delta T\le t\le T$,} \\
    0 & \text{else,}
  \end{cases}
\end{equation}
which allows for a smooth turn-on and turn-off of the laser field.  The
variables $T$ and $\Delta T$ denote the total interaction time and the
time of turn-on and turn-off.  After turn-on and before turn-off, the electric and
magnetic amplitudes of the oscillating electromagnetic fields are given
by
\begin{align}
 \vec {\hat E} & = k \vec {\hat A}_\mathrm{max}\,, \\
 \vec {\hat B} & = -\vec k \times \vec {\hat A}_\mathrm{max} \,.
\end{align}
For convenience, we will choose our coordinate system such that the
orthogonal vectors $\vec k$, $\vec {\hat B}$, and $\vec {\hat E}$ point
along the $x$, $y$, and $z$ directions.  In the following all vector
quantities will be projected in the directions of $\vec k$, $\vec {\hat
  E}$, and $\vec {\hat B}$ as indicated by the indices $k$, $E$, and
$B$, respectively.  For simplicity, we omit the index $E$ for the vector
potential, because the vector potential always points in the electric
field direction, that is, $\hat A(t)={\hat A}_E(t)$.

\subsection{The Pauli equation}
\label{sec:Pauli_eq}

The Pauli equation that governs the quantum motion of a spin-$1/2$
particle of mass $m$ and charge $q$ is given by
\begin{equation}
  \label{eq:pauli-equation}
  \I \dot \psi(\vec x,t) =  
  \frac{1}{2 m} \left( 
    -\I\nabla - \frac{q}{c} \vec A(\vec x,t)
  \right)^2 \psi(\vec x,t) - \frac{q}{2 m c} \vec\sigma \cdot \vec B(\vec x,t) \,
  \psi(\vec x,t)
\end{equation}
with the Pauli matrices
$\vec\sigma=\trans{(\sigma_1,\sigma_2,\sigma_3)}=\trans{(\sigma_k,\sigma_B,\sigma_E)}$
and $\vec B(\vec x,t) =\mbox{$\nabla\times \vec A(\vec x,t)$}$.  Taking
advantage of the sinusoidal spatial periodicity of the vector potential
$\vec A(\vec x,t)$, we expand the wave function
\begin{equation}
  \label{eq:pauli-wave-function}
  \psi(\vec x,t)=\sum_{n,\zeta} c_n^\zeta(t) \psi_n^\zeta(\vec x)
\end{equation}
into momentum eigenfunctions of the free Pauli Hamiltonian
\begin{equation}
  \label{eq:pauli-spinors}
  \psi^{\zeta}_n(\vec x) = 
  \sqrt{\frac{k}{2 \pi}} \chi^\zeta \e^{\I \vec p_n\cdot\vec x}
\end{equation}
with $\zeta\in\{\uparrow, \downarrow\}$, the two spinor basis functions
\begin{align}
  \chi^\uparrow & = 
  \begin{pmatrix}
    1 \\ 0
  \end{pmatrix}\,, &
  \chi^\downarrow & = 
  \begin{pmatrix}
    0 \\ 1
  \end{pmatrix}\,,
\end{align}
and the momentum $\vec p_n = \vec p + n \vec k$.  Inserting the ansatz
\eqref{eq:pauli-wave-function} into the Pauli equation
\eqref{eq:pauli-equation} and projecting onto the basis elements
\eqref{eq:pauli-spinors} from the left-hand side yields the Pauli
equation in momentum space
\begin{multline}
  \label{eq:pauli-equation-momentum-space}
  \I \dot{c}_n(t) = \\
  \frac{ (\vec p + n \vec k)^2}{2 m} c_n(t) 
  + \frac{q^2 \hat A(t)^2}{8 m c^2} \sin^2(\omega t) \left[c_{n-2}(t) + 2 c_n(t) + c_{n+2}(t)\right] \\
  - \frac{q \sin(\omega t)}{4 m c} \sum_{\delta n \in \{1, -1\}}\left[ 
    2 \hat A(t)p_E\,\mathds{1} - \I\delta n \vec{\hat k}\times\vec{\hat A}(t) \sigma_B
  \right] c_{n+\delta n}(t)\,,
\end{multline}
where the vectors $c_n(t)=\trans{(c_n^\uparrow(t),c_n^\downarrow(t))}$
and the $2\times2$ identity matrix $\mathds{1}$ have been introduced.

\subsection{The Dirac equation}
\label{sec:Dirac_eq}

The Dirac equation is a relativistic generalization of the
nonrelativistic Pauli equation. It is given by
\begin{equation}
   \label{eq:dirac-hamiltonian}
   \I \dot \psi(\vec x,t)  = 
   c \vec \alpha\cdot\left(
     -\I\nabla- \frac{q}{c} \vec A(\vec x, t)
   \right)  \, \psi(\vec x,t) + m c^2 \beta \, \psi(\vec x,t)
\end{equation}
with $\vec\alpha=\trans{(\alpha_1,\alpha_2,\alpha_3)}$ and $\beta$
denoting the Dirac matrices \cite{Thaller:1992:Dirac}.  We transform the Dirac equation
\eqref{eq:dirac-hamiltonian} into momentum space in close analogy to the
momentum space transformation of the Pauli equation
\eqref{eq:pauli-equation-momentum-space}.  For this purpose, the basis
elements \eqref{eq:pauli-spinors} of the wave function
\eqref{eq:pauli-wave-function} are replaced by
\begin{equation}
  \label{eq:dirac-momentum-eigen}
  \psi_n^{\gamma}(\vec x) = 
  \sqrt{\frac{k}{2 \pi}} u_n^\gamma \e^{\I \vec p_n \cdot\vec x}\,.
\end{equation}
The bi-spinors $u_n^\gamma$ with $\gamma\in\{+\uparrow, -\uparrow,
+\downarrow, -\downarrow\}$ are explicitly given by
\begin{subequations}
  \label{eq:dirac-bi-spinors}
  \begin{equation}
    u_n^{+\zeta} = \sqrt{\frac{\E_n + m c^2}{2 \E_n}}
    \begin{pmatrix}
      \chi^\zeta \\[0.5ex]
      \frac{\vec \sigma\cdot c \vec p_n}{\E_n + m c^2} \chi^\zeta
    \end{pmatrix}
  \end{equation}
  and 
  \begin{equation}
    u_n^{-\zeta} = \sqrt{\frac{\E_n + m c^2}{2 \E_n}}
    \begin{pmatrix}
      - \frac{\vec \sigma\cdot c \vec p_n}{\E_n + m c^2} \chi^\zeta \\[0.5ex]
      \chi^\zeta
    \end{pmatrix}\,.
  \end{equation}
\end{subequations}
The basis functions $\psi_n^{\pm\zeta}(\vec x)$ are simultaneously
eigenfunctions of the free time-independent Dirac equation with energy
eigenvalue $\pm\E_n$ defined as
\begin{equation}
  \label{eq:relativistic_enery_momentum_relation_modes}
  \E_n = \sqrt{m c^2 + c^2 \vec p_n^2}
\end{equation}
and eigenfunctions of the momentum operator with momentum eigenvalue
$\vec p_n$.  Furthermore, $\psi_n^{\pm\uparrow}(\vec x)$ and
$\psi_n^{\pm\downarrow}(\vec x)$ are eigenfunctions of the
Foldy-Wouthuysen spin operator along the $\vec{\hat E}$ direction with
eigenvalues $1/2$ and $-1/2$
\cite{Foldy:Wouthuysen:1950:nr_limit,deVries:1970:FW_Transformation,Schweber:2005:qft,Bauke:etal:2013:relativistic_hydrogen_spin}.

Inserting the ansatz \eqref{eq:pauli-wave-function} with the basis
elements \eqref{eq:dirac-bi-spinors} in the Dirac equation
\eqref{eq:pauli-equation} and projecting with these from the left-hand
side yields the Dirac equation in momentum space
\begin{equation}
  \label{eq:dirac-equation-momentum-space}
  \I \dot c_n(t) = 
  \E_n \beta c_n(t) - \frac{q \sin(\omega t)}{2}
  \left(L_{n,n-1} c_{n-1}(t) + L_{n,n+1} c_{n+1}(t) \right)\,, 
\end{equation}
with the vectors
\begin{equation}
  \label{eq:dirac_expansion_coefficient_vector}
  c_n(t)=
  \trans{(c_n^{+\uparrow}(t),c_n^{+\downarrow}(t),c_n^{-\uparrow}(t),c_n^{-\downarrow}(t))}
\end{equation}
and the coupling matrices $L_{n,n'}$ whose elements are defined as
\begin{equation}
  \label{eq:L_nm}
  L_{n,n'}^{\gamma,\gamma'} = 
  {u_n^{\gamma}}^\dagger \left( 
    \vec \alpha \cdot \vec {\hat A}(t) 
  \right) u_{n'}^{\gamma'}\,.
\end{equation}

\subsection{Numerical procedure}
\label{sec:numerical_procedure}

The absolute square values of the expansion coefficients $c_n^\zeta(t)$
in \eqref{eq:pauli-wave-function} represent the probability of finding
the electron in a particular free-particle quantum state.  The state
that is represented by $c_n^\zeta(t)$ has definite spin that is encoded
in the index $\zeta$ and also has a definite momentum $\vec p_n=\vec p +
n \vec k$.  Thus, the index $n$ counts the number of laser photon
momenta relative to the reference momentum $\vec p$.  For convenience,
we use the term ``mode $n$'' for these different electron momenta.

In numerical simulations, we start from an initial quantum state with
definite momentum $\vec p$ and spin-up polarization.  This means
\begin{equation}
   \label{eq:initial_condition}
   c_n^{\gamma}(0) =
   \begin{cases}
     1 & \text{if $n=0$ and $\gamma={+\uparrow}$ or
       $\gamma={\uparrow}$,} \\
     0 & \text{else.}
   \end{cases}
\end{equation}
The case $\gamma={+\uparrow}$ applies for the Dirac equation, while
$\gamma={\uparrow}$ is for the Pauli equation.  The initial electron
momentum $\vec p$, the laser intensity, and the laser frequency are
chosen to meet the nonrelativistic Bragg condition
(\ref{eq:p_k_nonrelativistic}) or its relativistic generalization
(\ref{eq:p_k_relativistic}) depending on whether the Pauli equation or
the Dirac equation is solved numerically.  The numerical solution of the
differential equations \eqref{eq:pauli-equation-momentum-space}
and~\eqref{eq:dirac-equation-momentum-space} is obtained by employing a
Crank-Nicholson scheme \cite{Numerical_Recipes:2002:Press}.
Equations~\eqref{eq:pauli-equation-momentum-space}
and~\eqref{eq:dirac-equation-momentum-space} couple an infinite number
of modes.  In numerical simulations, however, these systems are
truncated to a finite number of modes, $-n_\mathrm{max}\le n\le
n_\mathrm{max}$ with $n_\mathrm{max}$ large enough such that the
physical results are independent of $n_\mathrm{max}$.  The number of
included modes depends on the laser parameters and ranges typically from
one dozen to several dozens.  The duration of the turn-on and turn-off
phases $\Delta T$ is ten laser periods for all simulations presented in
this article unless another turn-on and turn-off time is indicated.

Note that our approach to solving the Dirac equation is tailored for
interactions with monochromatic laser fields.  More general approaches
include solving the Dirac equation via a Fourier transform
split-operator method \cite{Braun:Su:Grobe:1999:FFT_split_operator}, in
particular by employing geometric algebra
\cite{Mocken:2008:FFT_split_operator} and by making use of a graphics
processing unit \cite{Bauke:Keitel:2011:Accelerating}. Other methods
employ spherical harmonics as basis functions and Runge-Kutta
integration
\cite{Selsto:Lindroth:Bengtsson:2009:hydrogen_dirac_solution} or are
based on the method of characteristics
\cite{FillionGourdeau:Lorin:Bandrauk:2012:solution_dirac_equation}.

\section{Two-photon Kapitza-Dirac effect}
\label{sec:two_photon_KDE}

\subsection{Numerical results}
\label{sec:two_photon_KDE_numeric}

\begin{figure}
  \centering
  \includegraphics[scale=0.45]{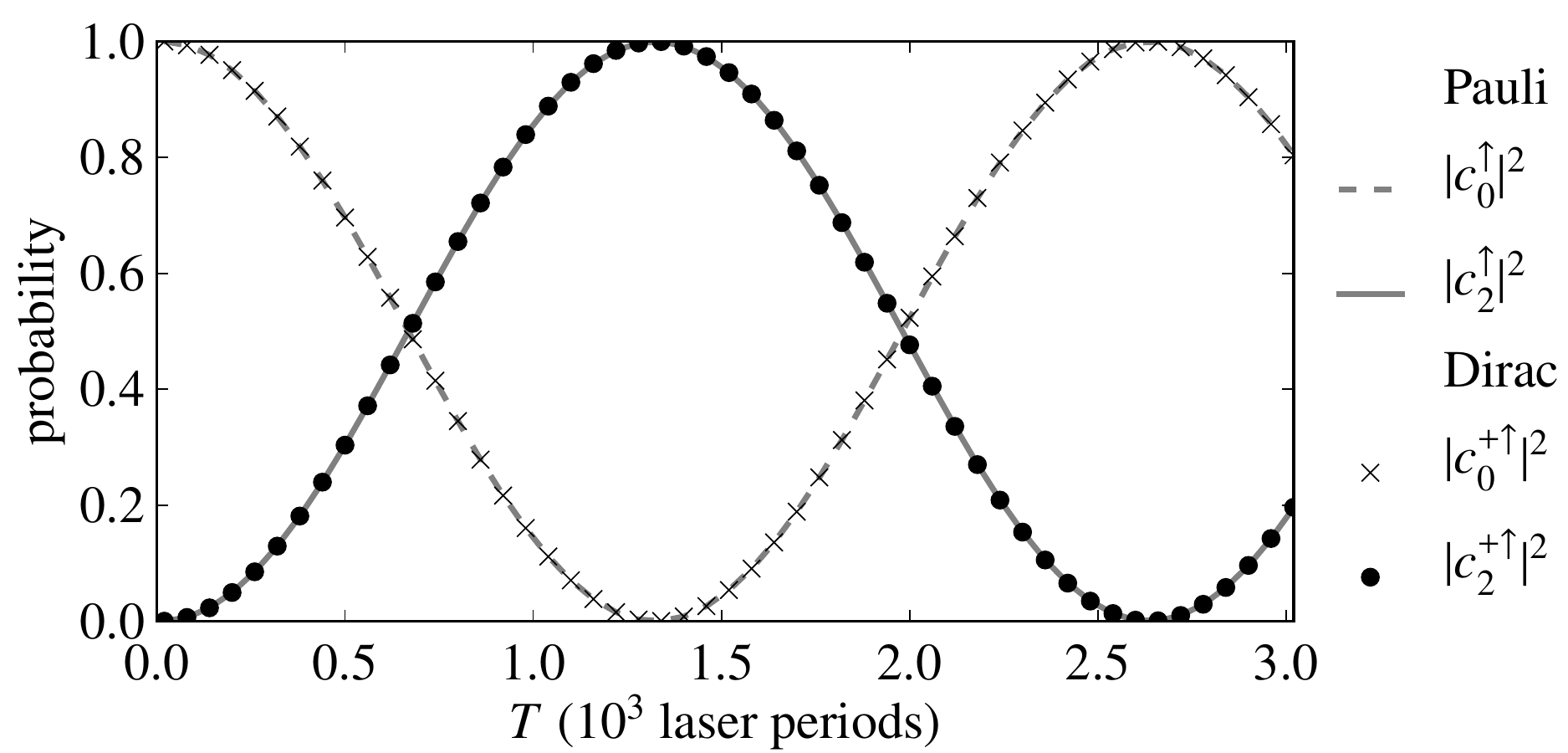}
  \caption{Occupation probabilities as a function of the total
    interaction time $T$ for the two-photon Kapitza-Dirac effect
    calculated by employing the Dirac equation (crosses and dots)
    \eqref{eq:dirac-equation-momentum-space} and the Pauli equation
    \eqref{eq:pauli-equation-momentum-space} (solid and dashed
    lines). The electron enters the laser field with a momentum
    component of $p_E=-12.5\,\text{keV}/c = 2.4\times10^{-2}mc$ along
    the laser electric field direction and interacts with the standing
    laser field of peak intensity $5\times
    10^{21}\,\textrm{W}/\textrm{cm}^2$ (corresponding to $e\hat
    A_\mathrm{max}=8.6\times10^{-3}mc^2$) for each laser beam and a
    photon energy of 12.5\,keV (corresponding to a laser wave length of
    $\lambda=0.1$\,nm). The probability that the electron is diffracted
    from mode~$0$ to mode~$2$ oscillates in Rabi cycles as a function of
    the total interaction time $T$.  The laser field is modulated via
    the envelope function \eqref{eq:turn_on_and_off} with a turn-on and
    turn-off time of ten laser periods.}
  \label{fig:original_kde}
\end{figure}

In the electron scattering dynamics as described by Kapitza and Dirac
\cite{Kapitza:Dirac:1933:effect} one photon is absorbed from the laser
field and one is emitted into the laser field, $n_r=1$ and $n_l=-1$ in
our notation.  For realizing this effect the parameters of the laser as
well as of the incident electron have to be chosen such that the quantum
dynamics is in the Bragg regime \cite{Batelaan:2000:Kapitza-Dirac}.  For
$n_r+n_l=0$ the nonrelativistic Bragg condition
(\ref{eq:p_k_nonrelativistic}) as well as the relativistic Bragg
condition (\ref{eq:p_k_relativistic}) require that the electron enters
the laser beam with $p_k=-k$.  For the numerical simulations shown in
Fig.~\ref{fig:original_kde} we choose $\vec p=-k\vec e_k+k\vec e_E$ with
$k=2.4\times10^{-2}mc$ and $e\hat A_\mathrm{max}=8.6\times10^{-3}mc^2$.

Figure~\ref{fig:original_kde} shows the quantum state after turn-off of
the laser for the two-photon Kapitza-Dirac effect for different total
interaction times $T$ as calculated by solving the Dirac and the Pauli
equation numerically.  As all parameters are in the nonrelativistic
regime the Dirac equation and the Pauli equation give qualitatively and
quantitatively the same results.  The quantum dynamics exhibits the
well-known Rabi oscillations of the diffraction probability
\cite{Batelaan:2000:Kapitza-Dirac} from mode~$0$ with momentum $\vec p =
-k\vec e_k+k\vec e_E$ to mode~$2$ with momentum $\vec p =+k\vec
e_k+k\vec e_E$, in the form
\begin{subequations}
  \label{eq:2-photon_KDE_rabioscillation}
  \begin{align}
    \label{eq:2-photon_KDE_rabioscillation_cos}
    |c_{0}^{+\uparrow}(T)|^2 & = 
    \cos^2 \left(\frac{\Omega_R T}{2} \right) \,,\\
    \label{eq:2-photon_KDE_rabioscillation_sin}
    |c_{2}^{+\uparrow}(T)|^2 &=  
    \sin^2 \left(\frac{\Omega_R T}{2} \right)\,.
  \end{align}
\end{subequations}
Here, $\Omega_R$ denotes the Rabi frequency. The occupation probability
of all other modes is vanishingly small, in particular,
$|c_2^{+\downarrow}(T)|^2<10^{-6}$.  This means that no spin flip occurs
during two-photon Kapitza-Dirac scattering.  From a naive point of view
the vanishing spin-flip probability might be surprising, because one
might expect a precession of the electron spin in the magnetic field of
the external laser field. The question of electron spin precession has
already been investigated based on nonrelativistic classical equations
of motion \cite{Freimund:Batelaan:2003:Stern-Gerlach} but no significant
spin effects could be found.  In Sec.~\ref{sec:two_photon_spin_flip},
however, we will show that spin-flips are possible in the two-photon
Kapitza-Dirac effect for certain relativistic parameter settings.

\subsection{Perturbation theory}
\label{eq:2-photon_KDE_perturbation_theory}

In the following we will complement our numerical findings with
analytical results obtained via time-dependent perturbation
theory. Time-dependent perturbation theory for the Dirac
equation~\eqref{eq:dirac-equation-momentum-space} will allow us to
calculate analytical expressions for the Rabi frequency and to derive
conditions that permit spin-flip dynamics in the two-photon
Kapitza-Dirac effect. The perturbative solution also allows for
deduction of a rotation of the electron spin during diffraction. As the
initial condition is given by \eqref{eq:initial_condition} and the
electron momentum is changed by two photon momenta our aim is to
approximate the time evolution operator $U_{2,0}(t, 0)$ that maps
$c_0(0)$ to $c_2(t)$, viz.,
\begin{equation}
  \label{eq:c_2_propagation}
  c_2(t) = U_{2,0}(t, 0) c_0(0)\,.
\end{equation}

Since the Dirac equation~\eqref{eq:dirac-equation-momentum-space}
couples next neighboring modes only, the lowest non-vanishing
contribution to $U_{2,0}(t, 0)$ is of second order in time-dependent
perturbation theory.  The general second order propagator for a
time-dependent Hamiltonian $H(t)$ reads
\cite{Joachain:etal:2011:Atoms_in_Intense_Laser_Fields}
\begin{multline}
  \label{eq:second-order-perturbation-theory}
  U_\mathrm{nd}(t,0) = \\
  \frac{1}{\I^2} 
  \int_{0}^{t} \D t_2 
  \int_{0}^{t_2} \D t_1 
  U_0(t,t_2)V(t_2) U_0(t_2,t_1) V(t_1) U_0(t_1,0)\,.
\end{multline}
The symbol $U_0(t,0)$ denotes the free propagator
\begin{math}
  U_0(t,0)=\e^{- \I {H}_0 t}\,,
\end{math}
with the time-independent field-free Hamiltonian
\begin{equation}
  H_0 =
  \begin{pmatrix}
    \ddots & & & & \\
    & H_{0;1,1} & & &\\
    & & H_{0;0,0} & & \\
    & & & H_{0;-1,-1} & \\
    & & & & \ddots 
  \end{pmatrix}
\end{equation}
with $H_{0;a,b} = \mathcal{E}_a\beta\delta_{a,b}$ and the time-dependent
interaction Hamiltonian \mbox{$V(t) = H(t) - H_0$}.  For the Dirac
equation in momentum space \eqref{eq:dirac-equation-momentum-space}, the
free propagator reads explicitly
\begin{equation}
  \label{eq:free-propagator}
  U_{0;a,b}(t,0) =
  \e^{-\I \E_a \beta t}\delta_{a,b} 
  =
  \begin{pmatrix}
    \exp(- \I \E_a t) \mathds{1} & 0 \\
    0 & \exp(\I \E_a t) \mathds{1}
  \end{pmatrix}\delta_{a,b}\,.
\end{equation}
The corresponding interaction Hamiltonian reads
\begin{align}
  \label{eq:interaction-hamiltonian}
  V_{a,b}(t) = 
  - \frac{q \sin(\omega t)}{2} \left( 
    L_{a,a-1} \delta_{a,b+1} + L_{a,a+1} \delta_{a,b-1}
  \right)\,.
\end{align}
Inserting these expressions into
\eqref{eq:second-order-perturbation-theory} yields
\begin{multline}
  \label{eq:dirac_propagator_second_order_perturbation_theory_explicit}
  U_{\mathrm{nd};2,0}(t,0) = 
  -
  \int_{0}^{t} \D t_2  
  \int_{0}^{t_2} \D t_1 \e^{- \I \E_2 \beta (t-t_2)}
  \frac{q \sin(\omega t_2)}{2} L_{2,1} \\
  \times \e^{- \I \E_1 \beta (t_2-t_1)} 
  \frac{q \sin(\omega t_1)}{2} L_{1,0} 
  \e^{- \I \E_0 \beta t_1} \,.
\end{multline}

In order to ease notations, it will be useful to split the $4\times 4$
matrices $L_{n,n'}$ and $U_{\mathrm{nd};2,0}(t,0)$ into blocks of
$2\times 2$ matrices, viz.,
\begin{equation}
  \label{eq:spin_interaction_matices}
  L_{n,n'} =
  \begin{pmatrix}
    L_{n,n'}^{++} & L_{n,n'}^{+-} \\[0.75ex]
    L_{n,n'}^{-+} & L_{n,n'}^{--}
  \end{pmatrix}
  \quad\text{with}\quad
  L_{n,n'}^{ab} =
  \begin{pmatrix}
    L_{n,n'}^{a\uparrow, b\uparrow} & L_{n,n'}^{a\uparrow, b\downarrow} \\[0.75ex]
    L_{n,n'}^{a\downarrow, b\uparrow} & L_{n,n'}^{a\downarrow, b\downarrow}
  \end{pmatrix}
\end{equation}
and
\begin{equation}
  U_{\mathrm{nd};2,0}(t,0) =
  \begin{pmatrix}
    U_{\mathrm{nd};2,0}^{++}(t,0) &  U_{\mathrm{nd};2,0}^{+-}(t,0) \\[0.75ex]
    U_{\mathrm{nd};2,0}^{-+}(t,0) &  U_{\mathrm{nd};2,0}^{--}(t,0) 
  \end{pmatrix}\,.
\end{equation}
Explicit expressions for the matrices $L_{n,n'}^{ab}$ are given in the
Appendix~\ref{sec:appendix_spin-matrices}. With these definitions the
sub-propagator $U_{\mathrm{nd};2,0}^{++}(t,0)$ in the space of
positive-energy free-particle states reads for times after the turn-on
phase and before the turn-off phase ($\Delta T=0$ and $0<t<T$)
\begin{widetext}
  \begin{equation}
    \label{eq:dirac_propagator_second_order_perturbation_theory_expanded}
    U_{\mathrm{nd};2,0}^{++}(t,0) = 
    - \frac{q^2}{4} L_{2,1}^{++} L_{1,0}^{++}
    \int_{0}^{t} \D t_2  
    \int_{0}^{t_2} \D t_1 \sin(\omega t_2) \sin(\omega t_1) 
    v^+(t,t_2,t_1) 
    - \frac{q^2}{4} L_{2,1}^{+-} L_{1,0}^{-+} 
    \int_{0}^{t} \D t_2  
    \int_{0}^{t_2} \D t_1 \sin(\omega t_2) \sin(\omega t_1) 
    v^-(t,t_2,t_1)\,,
  \end{equation}
  where we have introduced the two complex phases
  \begin{align}
    \label{eq:dirac_second_order_perturbation_theory_phase_1}
    v^+(t,t_2,t_1) & = \exp\left[
      - \I \left(
        \E_2 t + \Delta\E_{1,2}^{++} t_2 + \Delta\E_{0,1}^{++} t_1
      \right) 
    \right] \,,\\
    \label{eq:dirac_second_perturbation_theory_phase_2}
    v^-(t,t_2,t_1) & = \exp\left[
      - \I \left( 
        \E_2 t + \Delta\E_{1,2}^{-+} t_2  + \Delta\E_{0,1}^{+-} t_1
      \right) 
    \right]\,.
  \end{align}
  Here, $\Delta \E_{n,n'}^{ab}$ is an abbreviation for the energy difference
  \begin{math}
    \Delta \E_{n,n'}^{ab} = \sign(a) \E_n - \sign(b) \E_{n'}\,,
  \end{math}
  where the signum of the upper indices is $\sign(+)=1$ and $\sign(-)=-1$.
  Performing the first integral in
  \eqref{eq:dirac_propagator_second_order_perturbation_theory_expanded} we
  find
  \begin{multline}
    \label{eq:int_t2}
    \int_{0}^{t} \D t_2 \int_{0}^{t_2} \D t_1 \sin(\omega t_2)
    \sin(\omega t_1)
    v^+(t,t_2,t_1) = \\
    -\frac{1}{4} \e^{-\I\E_2t} \int_{0}^{t} \D t_2 \Bigg[
    \frac{\I}{\Delta\E^{++}_{0,1}+\omega}\left(
      \e^{-\I(\Delta\E^{++}_{0,2}+2\omega)t_2} - 
      \e^{-\I(\Delta\E^{++}_{1,2}+\omega)t_2}
    \right)-
    \frac{\I}{\Delta\E^{++}_{0,1}-\omega}\left( 
      \e^{-\I\Delta\E^{++}_{0,2}t_2} - 
      \e^{-\I(\Delta\E^{++}_{1,2}+\omega)t_2}
    \right) \\ -
    \frac{\I}{\Delta\E^{++}_{0,1}+\omega}\left(
      \e^{-\I\Delta\E^{++}_{0,2}t_2} - 
      \e^{-\I(\Delta\E^{++}_{1,2}-\omega)t_2}
    \right)+
    \frac{\I}{\Delta\E^{++}_{0,1}-\omega}\left( 
      \e^{-\I(\Delta\E^{++}_{0,2}-2\omega)t_2} - 
      \e^{-\I(\Delta\E^{++}_{1,2}-\omega)t_2}
    \right) 
    \Bigg]\,.
  \end{multline}
\end{widetext}
The integral \eqref{eq:int_t2} may show linear or oscillating behavior
depending on the laser parameters and the initial electron momentum
$\vec p$.  For transitions from mode $0$ to mode $2$ the absolute value
of the coefficient $c_2(t)$ must grow linearly in $t$ within
perturbation theory. The expression \eqref{eq:int_t2} and therefore the
propagator
\eqref{eq:dirac_propagator_second_order_perturbation_theory_expanded}
feature terms growing linearly in $t$ if and only if at least one of the
exponents on the right-hand side of \eqref{eq:int_t2} is zero.  Taking
into account $|\Delta\E^{++}_{n,n'}|<|n-n'|\omega$, this leads us to the
unique resonance condition $\Delta \E_{0,2}^{++}=0$ which is equivalent
to the classical energy-momentum conservation conditions
\eqref{eq:momentum-conservation} and \eqref{eq:energy-conservation} with
$\vec p_\mathrm{in}=\vec p$ and $n_r=1$ and $n_l=-1$.  The second
integral in
\eqref{eq:dirac_propagator_second_order_perturbation_theory_expanded}
leads to the same resonance condition.  Thus, on resonance the
propagator $U_{\mathrm{nd};2,0}^{++}(t,0)$ reads in leading order in~$t$
\begin{equation}
  \label{eq:dirac-second_order_perturbation-theory-result}
  U_{\mathrm{nd};2,0}^{++}(t,0) = 
  - \frac{\I q^2 t}{16} \e^{-\I\E_2 t}\left(
    l^+ L_{2,1}^{++} L_{1,0}^{++} + l^- L_{2,1}^{+-}L_{1,0}^{-+} 
  \right) 
\end{equation}
with the coefficients 
\begin{subequations}
  \begin{align}
    l^+ & = 
    \frac{1}{\Delta \E^{++}_{0,1} - \omega} +  
    \frac{1}{\Delta \E^{++}_{0,1} + \omega} \,,\\
    l^- & = 
    \frac{1}{\Delta \E^{+-}_{0,1} - \omega} +  
    \frac{1}{\Delta \E^{+-}_{0,1} + \omega} \,.
  \end{align}
\end{subequations}

\subsection{The relativistic Rabi frequency}
\label{sec:2-photon_KDE_Rabifrequency}

Employing the explicit form of the matrices $L_{n,n'}^{ab}$ in
Eqs.~\eqref{eq:dirac-energy-keep-matrix} and
\eqref{eq:dirac-energy-change-matrix} and the propagator
\eqref{eq:dirac-second_order_perturbation-theory-result} we calculate
the diffraction probability $|c_2^{+\uparrow}(t)|^2 +
|c_2^{+\downarrow}(t)|^2$ which equals with
$c_n^{+}(t)=\trans{(c_n^{+\uparrow}(t), c_n^{+\downarrow}(t))}$
\begin{multline}
  \label{eq:norm_c_2_t}
  |c_2^{+\uparrow}(t)|^2 + |c_2^{+\downarrow}(t)|^2 =
  c_2^{+}(t)^\dagger c_2^{+}(t) \\=
  c_2^{+}(0)^\dagger
  U_{\mathrm{nd};2,0}^{++}(t,0)^\dagger U_{\mathrm{nd};2,0}^{++}(t,0)
  c_2^{+}(0)\,.
\end{multline}
Expanding $U_{\mathrm{nd};2,0}^{++}(t,0)$ from
Eq.~\eqref{eq:dirac-second_order_perturbation-theory-result} in the
momenta $k$, $p_B$, and $p_E$ yields
\begin{multline}
  \label{eq:taylored_2_photon_KDE_propagator}
  U_{\mathrm{nd};2,0}^{++}(t,0) =  
  - \frac{\I q^2 t}{16} 
  \frac{\hat A_\mathrm{max}^2}{m c^2} \frac{\e^{-\I\E_2 t}}{m^2 c^2} \left[
    \left(m^2 c^2 - \frac{p_B^2}{2} - \frac{5p_E^2}{2} 
    \right) \mathds{1} \right. \\
  + \left.
    \frac{\I k p_B}{2} \sigma_E +
    \frac{\I 3k p_E}{2} \sigma_B 
  \right] 
  + \mathcal{O}(k^3,p_E^3,p_B^3)\,.
\end{multline}
Taking advantage of the normalization of the initial state, i.\,e.,
$|c_0^{+\uparrow}(0)|^2+|c_0^{+\downarrow}(0)|^2=1$, we finally get with
\eqref{eq:norm_c_2_t} and \eqref{eq:taylored_2_photon_KDE_propagator}
\begin{equation}
  \label{eq:quadratic_grwoth}
  |c_2^{+\uparrow}(t)|^2 + |c_2^{+\downarrow}(t)|^2 =
  \frac{t^2}{4}\left(
    \frac{q^2\hat E^2}{8k^2mc^2}
  \right)^2
\end{equation}
in leading order in $k$, $p_B$, and $p_E$.  For times of the order of
${8k^2mc^2}/({q^2\hat E^2})$ or larger we leave the domain where our
time-dependent perturbation theory is valid.  Thus, the result
\eqref{eq:quadratic_grwoth} is correct for times much shorter than
${8k^2mc^2}/({q^2\hat E^2})$ only.  Making the ansatz
\begin{subequations}
  \label{eq:2-photon_KDE_rabioscillation-ansatz}
  \begin{align}
    \label{eq:2-photon_KDE_rabioscillation-ansatz_0}
    |c_{0}^{+\uparrow}(t)|^2 & =
    \cos^2 \left(\frac{\Omega_R t}{2} \right) \,,\\
    \label{eq:2-photon_KDE_rabioscillation-ansatz_2}
    |c_2^{+\uparrow}(t)|^2 + |c_2^{+\downarrow}(t)|^2 & = \sin^2 \left(\frac{\Omega_R t}{2}
    \right) 
  \end{align}
\end{subequations}
for the long-time behavior, expanding this ansatz for short times, and
comparing it to \eqref{eq:quadratic_grwoth} gives the known Rabi
frequency of the two-photon Kapitza Dirac
effect~\cite{Batelaan:2000:Kapitza-Dirac,Batelaan:2007:Illuminating_Kapitza-Dirac}
\begin{equation}
  \label{eq:Rabi_freq_nr}
  \Omega_{R,2} = \frac{q^2\hat E^2}{8k^2mc^2}\,.
\end{equation}

\begin{figure}
  \centering
  \includegraphics[scale=0.45]{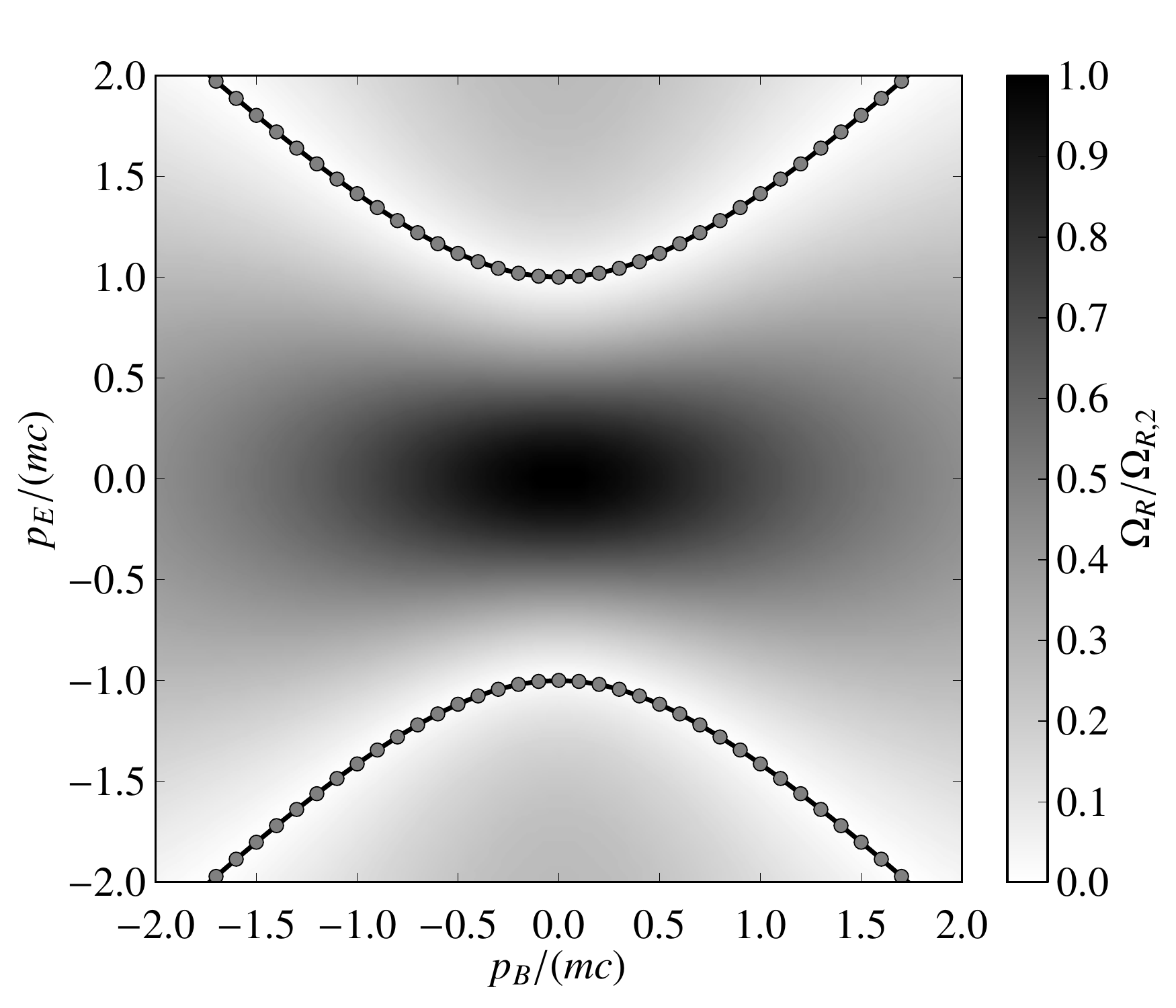}
  \caption{The relativistic Rabi frequency $\Omega_R$ of the two-photon
    Kapitza-Dirac effect normalized to the nonrelativistic Rabi
    frequency $\Omega_{R,2}$ in Eq.~\eqref{eq:Rabi_freq_nr}. The applied
    parameters correspond
    to 
    a photon energy of 12.5\,keV. The gray filled circles indicate
    momenta for which the condition
    \eqref{eq:propagator_spin-flip_condition} is fulfilled and full
    spin flips can be observed. These momenta can be approximated by
    Eq.~\eqref{eq:relativistic_spin-flip_condition} (solid black line).}
  \label{fig:2-photon_KDE-spin-flip_condition}
\end{figure}

The expression \eqref{eq:Rabi_freq_nr} for the Rabi frequency which is
based on the expansion \eqref{eq:taylored_2_photon_KDE_propagator} is
valid only in the nonrelativistic domain.  The Rabi frequency for
relativistic momenta $k$, $p_B$, and $p_E$ can be calculated, however,
with the help of computer algebra and numerical methods in a similar
fashion as \eqref{eq:Rabi_freq_nr} by evaluating \eqref{eq:norm_c_2_t}
and employing the fully relativistic propagator
\eqref{eq:dirac-second_order_perturbation-theory-result}.  The
relativistic Rabi frequency is shown
Fig.~\ref{fig:2-photon_KDE-spin-flip_condition}.  Generally the
relativistic Rabi frequency is lower than the nonrelativistic
result~\eqref{eq:Rabi_freq_nr}.  For the parameters as applied in the
setup of Fig.~\ref{fig:original_kde} the theoretical relativistic Rabi
frequency is $\Omega_R=7.241\times10^{15}\,\mathrm{Hz}$, while
numerically we obtain $\Omega_R=7.237\times10^{15}\,\mathrm{Hz}$, which
is in a fair agreement with the theoretical prediction.

\begin{figure}
  \centering
  \includegraphics[scale=0.45]{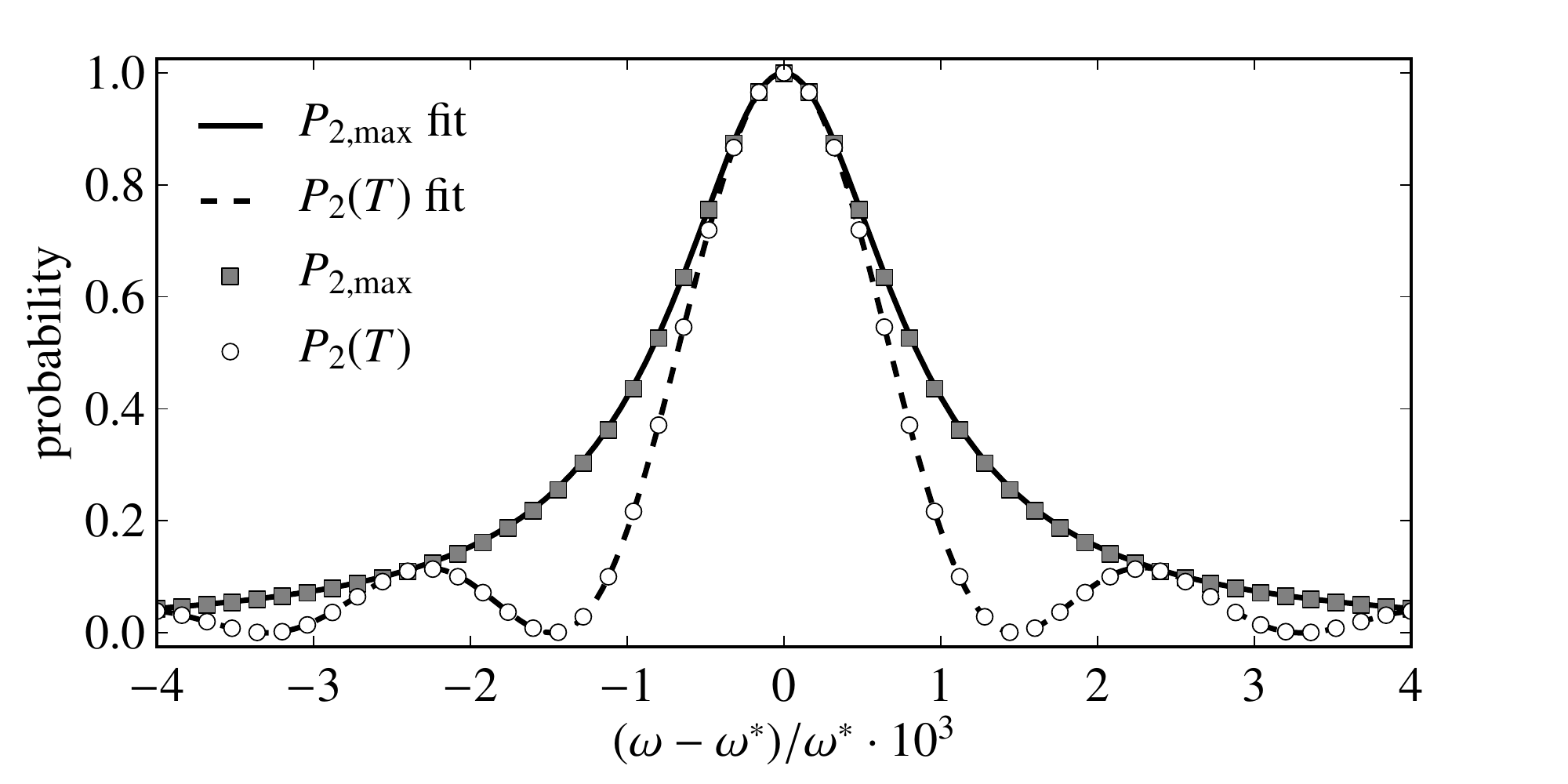}
  \caption{The off-resonant scattering probability
    $P_2(T)=|c_2^{+\uparrow}(T)|^2 + |c_2^{+\downarrow}(T)|^2$ at time
    $T=\pi/\Omega_R$ (half a Rabi cycle on resonance) and maximal
    scattering probability maximized over the total interaction time
    $P_{2,\mathrm{max}}=\max_{T'} P_2(T')$ as functions of the relative
    detuning \mbox{$(\omega-\omega^*)/\omega^*$}. Dots and squares
    represent numerical data while lines are obtained via a fit to
    \eqref{eq:2_off_resonant_scattering_prob} and
    \eqref{eq:generalized-detuning-rabi-frequency-relation} yielding
    $b=29.2$ for the fitting parameter.}
  \label{fig:resonance_peak}
\end{figure}

Let $k^*$ denote an electron momentum and a photon momentum that fulfill
the Bragg condition \eqref{eq:p_k_relativistic}. When the Bragg
condition \eqref{eq:p_k_relativistic} is not exactly fulfilled by
shifting the photon momentum $k$ from $k^*$, electrons scatter
off-resonantly, leading to a modification of the Rabi frequency and a
reduction of the maximal scattering probability.  In the resonant
two-photon Kapitza-Dirac effect only modes~0 and~2 are populated.  Thus,
the quantum system behaves similarly to an effective two-level
system. This leads us to the off-resonant generalization
\begin{equation}
  \label{eq:2_off_resonant_scattering_prob}
  |c_2^{+\uparrow}(T)|^2 + |c_2^{+\downarrow}(T)|^2 =
  \frac{\Omega_R^2}{\Omega^2} \sin^2\left(\frac{\Omega T}{2}\right)
\end{equation}
of \eqref{eq:2-photon_KDE_rabioscillation_sin} with the off-resonant
Rabi frequency  \cite{scully:zubairy:1997:quantum_optics}
\begin{equation}
  \label{eq:generalized-detuning-rabi-frequency-relation}
  \Omega=\sqrt{\Omega_R^2 + \frac{(\omega-\omega^*)^2}{b^2}}\,,
\end{equation}
$\omega=kc$, $\omega^*=k^*c$, and the parameter $b$ that accounts for
the fact that the two-photon Kapitza-Dirac effect is not a pure
two-level system.  Numerical simulations indicate that the parameter $b$
varies with the laser frequency and the electron momentum and has the
value $b=29.2$ for $p_B=0$ and $p_E = 1.00012 mc$ and $e\hat
A_\mathrm{max}=2.2\times10^{-2}mc^2$. The off-resonant scattering
probability $P_2(T)=|c_2^{+\uparrow}(T)|^2 + |c_2^{+\downarrow}(T)|^2$
at time $T=\pi/\Omega_R$ (half a Rabi cycle on resonance) and maximal
scattering probability maximized over the total interaction time
$P_{2,\mathrm{max}}=\max_{T'} P_2(T')$ obtained via numerical
simulations are shown in Fig.~\ref{fig:resonance_peak} as functions of
the relative detuning \mbox{$(\omega-\omega^*)/\omega^*$}.  These
numerical results can be fitted to
\eqref{eq:2_off_resonant_scattering_prob} and
\eqref{eq:generalized-detuning-rabi-frequency-relation}, respectively,
leading to the value $b=29.2$.

\subsection{Spin flips}
\label{sec:two_photon_spin_flip}

The propagator \eqref{eq:taylored_2_photon_KDE_propagator} features
spin-preserving terms (proportional to $\mathds{1}$) and spin-flipping
terms (proportional to $\sigma_E$ and $\sigma_B$).  If the condition 
\begin{equation}
   \label{eq:non-relativistic_spin-flip_condition}
   m^2 c^2 - \frac{1}{2} p_B^2 - \frac{5}{2} p_E^2 = 0
\end{equation}
is met, the propagator \eqref{eq:taylored_2_photon_KDE_propagator}
predicts that spin-preserving transitions are totally suppressed; thus,
a spin-flipping dynamics may become observable in the two-photon
Kapitza-Dirac effect.  However, this condition corresponds to an ellipse
with major axis $p_B = \sqrt{2} m c$ and minor axis $p_E = \sqrt{2/5} m
c$ and, therefore, we are beyond the validity of the nonrelativistic
propagator \eqref{eq:taylored_2_photon_KDE_propagator}.  Although
Eq.~\eqref{eq:non-relativistic_spin-flip_condition} is not a valid
condition for spin-flipping transitions its derivation gives us a hint
as to how to calculate the proper condition.  In analogy to
\eqref{eq:taylored_2_photon_KDE_propagator} one can expand the
relativistic propagator
\eqref{eq:dirac-second_order_perturbation-theory-result} as a
superposition of the matrices $\mathds{1}$, $\sigma_k$, $\sigma_B$, and
$\sigma_E$. With the help of the Frobenius inner product one can write
\begin{multline}
  \label{eq:propagator_decomposition}
  U_{\mathrm{nd};2,0}^{++}(t,0) =
  \tr\left(\mathds{1}^\dagger U_{\mathrm{nd};2,0}^{++}(t,0)\right) \tfrac{1}{2}\mathds{1} +
  \tr\left(\sigma_k^\dagger U_{\mathrm{nd};2,0}^{++}(t,0)\right) \tfrac{1}{2}\sigma_k \\+
  \tr\left(\sigma_B^\dagger U_{\mathrm{nd};2,0}^{++}(t,0)\right) \tfrac{1}{2}\sigma_B +
  \tr\left(\sigma_E^\dagger U_{\mathrm{nd};2,0}^{++}(t,0)\right) \tfrac{1}{2}\sigma_E \,.
\end{multline}
Thus, spin-preserving transitions are expected to be suppressed for
\begin{equation}
  \label{eq:propagator_spin-flip_condition}
  \tr\left(\mathds{1}^\dagger U_{\mathrm{nd};2,0}^{++}(t,0)\right)=0\,.
\end{equation}
Again this condition can be evaluated with the help of computer algebra
and numerical methods. The result is indicated in
Fig.~\ref{fig:2-photon_KDE-spin-flip_condition} by gray filled circles.
These points lie approximately on the hyperbola (black solid line in
Fig.~\ref{fig:2-photon_KDE-spin-flip_condition})
\begin{equation}
  \label{eq:relativistic_spin-flip_condition}
  m^2 c^2 + p_B^2 - p_E^2 = 0 \,,
\end{equation}
which can be seen as the fully relativistic version of the
condition~\eqref{eq:non-relativistic_spin-flip_condition}.  

\begin{figure}
  \centering
  \includegraphics[scale=0.45]{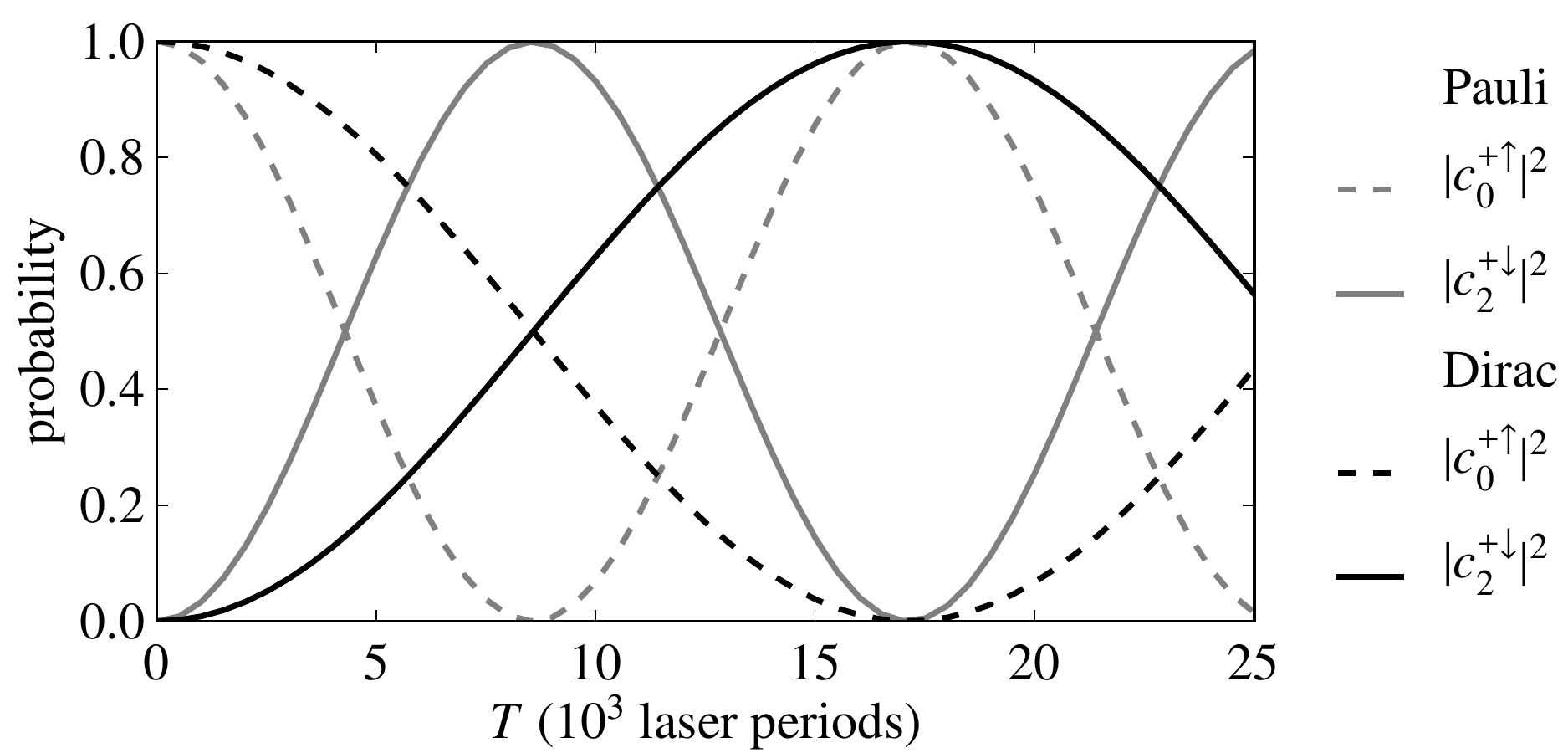}
  \caption{The two-photon Kapitza-Dirac effect with electron momenta
    chosen such that the spin-flip condition is fulfilled simulated by
    employing the Dirac equation
    \eqref{eq:dirac-equation-momentum-space} (black solid and dashed
    lines) and the Pauli equation
    \eqref{eq:pauli-equation-momentum-space} (gray solid and dashed
    lines).  The electron interacts with a standing laser field of peak
    intensity $3\times 10^{22}\,\mathrm{W}/\mathrm{cm}^2$ (corresponding
    to $e\hat A_\mathrm{max}=2.2\times10^{-2}mc^2$) for each beam and
    photon energy of 12.5\,keV. The electron is diffracted to mode~2, as
    in Fig.~\ref{fig:original_kde}.  However, because the initial
    electron momentum ($p_B=0$, $p_E = 1.00012 mc$) fulfills the
    spin-flip condition the electron spin changes its spin orientation
    during the diffraction process.}
  \label{fig:spinflip_2-photon_kde}
\end{figure}

\begin{figure}
  \centering
  \includegraphics[scale=0.45]{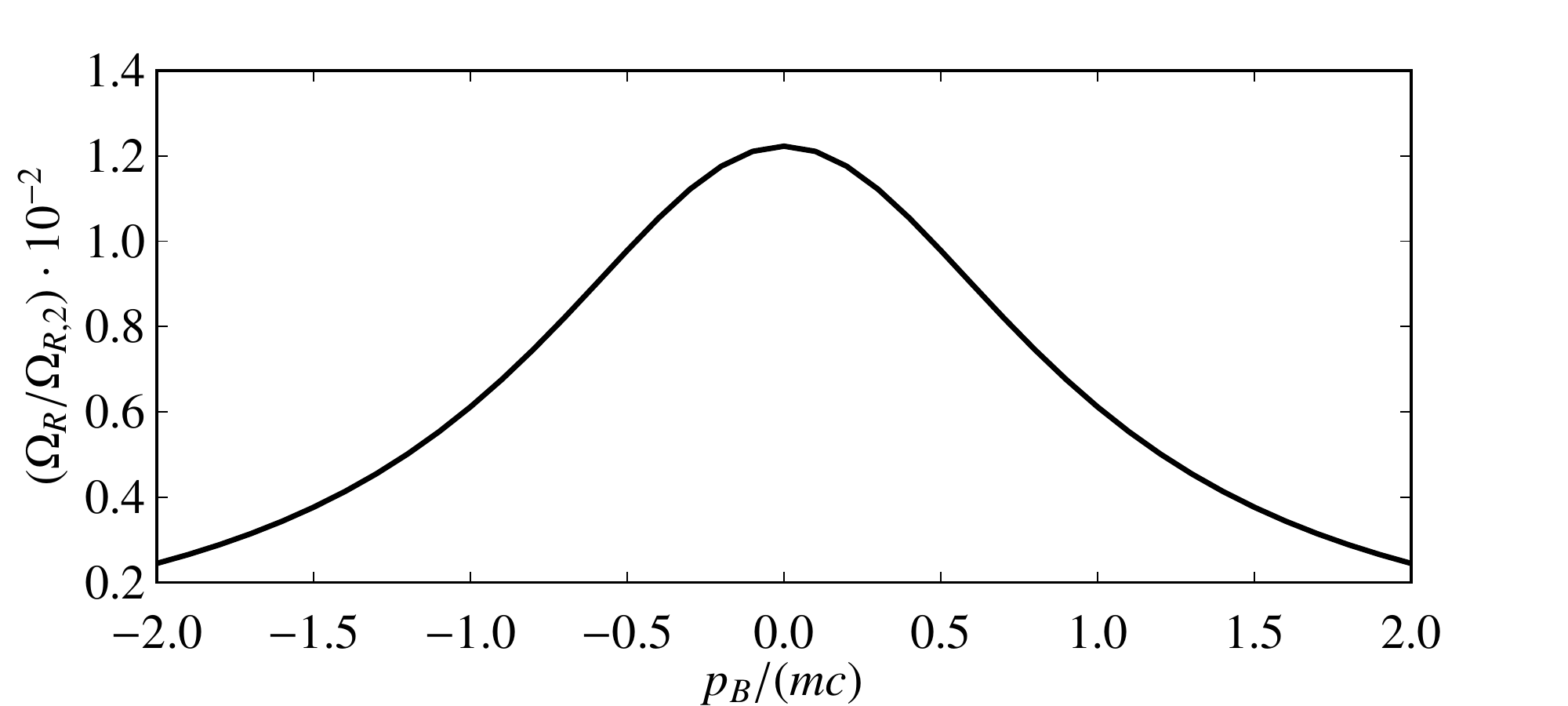}
  \caption{Relativistic Rabi frequency of the two-photon Kapitza-Dirac
    effect at momenta for which full spin flips are possible (that means
    along the gray filled circles in
    Fig.~\ref{fig:2-photon_KDE-spin-flip_condition}).}
  \label{fig:spinflip_Rabi-frequency}
\end{figure}

Employing time-dependent perturbation theory for the Pauli equation one
can show that $m^2c^2-p_E^2=0$ is the corresponding condition for
spin flips in the two-photon Kapitza-Dirac effect under the dynamics of
the Pauli equation.  Thus, the parameter setting $p_B=0$ and $p_E =mc$
fulfills the conditions for spin-flipping electron scattering under the
dynamics of the Dirac equation as well as under the Pauli equation, as
demonstrated in Fig.~\ref{fig:spinflip_2-photon_kde}. In contrast to the
setting in Fig.~\ref{fig:original_kde}, the Pauli equation and the Dirac
equation yield different Rabi frequencies due to the relativistic
electron momenta, for which the Pauli equation is actually not
applicable.  The hyperbola \eqref{eq:relativistic_spin-flip_condition}
lies approximately in a local minimum of the Rabi frequency (see
Fig.~\ref{fig:2-photon_KDE-spin-flip_condition}).  The Rabi frequency
along the gray filled dots in
Fig.~\ref{fig:2-photon_KDE-spin-flip_condition} is shown in
Fig.~\ref{fig:spinflip_Rabi-frequency}.  For photon momenta $k$ that are
small compared to $mc$ the Rabi frequency drops down by two orders of
magnitude compared to $\Omega_{R,2}$, which is a consequence of the
relative strengths of the spin-preserving and the spin-flipping terms
in~\eqref{eq:taylored_2_photon_KDE_propagator}.

Because the basis functions \eqref{eq:dirac-momentum-eigen} are
eigenfunctions of the Foldy-Wouthuysen spin operator in the laser's
polarization direction the coefficients $c_2^{+\uparrow}(t)$ and
$c_2^{+\downarrow}(t)$ give immediately the spin of the diffracted beam
if one chooses the electric field direction as quantization axis.
Perturbation theory allows us also to calculate the orientation of the
spin in the diffracted part of the electron beam.  More precisely, we
ask the question: What is the expectation value of the spin $\vec
s_\mathrm{out}$ in the scattered beam if the beam of incident electrons
has spin $\vec s_\mathrm{in}$?

The electron spin is determined by applying the Foldy-Wouthuysen spin
operator to the quantum state.  In our momentum-space representation the
expectation value of the Foldy-Wouthuysen spin operator is 
\begin{equation}
  \label{eq:FW}
  \vec s = \frac{1}{2}\sum_n c_n(t)^\dagger\vec\Sigma c_n(t)\,,
\end{equation}
with $\vec\Sigma=\trans{(\Sigma_1,\Sigma_2,\Sigma_3)}$ and 
\begin{equation}
  \Sigma_i =
  \begin{pmatrix}
    \sigma_i & 0 \\
    0 & \sigma_i
  \end{pmatrix}\,.
\end{equation}
The incident electron beam has well-defined momentum but may have
arbitrary spin orientation; thus it can be expressed as a superposition
of two positive-energy states with the same momentum.  Introducing the
Bloch angles $\theta$ and $\phi$ we may write the initial state as
\begin{equation}
  c_0(0) =
  \begin{pmatrix}
     c_0^{+\uparrow}(0) \\
     c_0^{+\downarrow}(0) \\
     c_0^{-\uparrow}(0) \\
     c_0^{-\downarrow}(0) 
  \end{pmatrix}
  =
  \begin{pmatrix}
     \cos(\theta/2) \\
     \sin(\theta/2) \e^{\I\phi} \\
     0 \\
     0
  \end{pmatrix}
\end{equation}
and $c_n(0)=\trans{(0,0,0,0)}$ for all $n\ne 0$.  The spin expectation
value of the initial quantum state is
\begin{equation}
  \vec s_\mathrm{in} =
  \frac{1}{2}
  \begin{pmatrix} 
    \sin(\theta)\cos(\phi) \\ 
    \sin(\theta)\sin(\phi) \\
    \cos(\theta) 
  \end{pmatrix}\,.
\end{equation}
The spin expectation value of the diffracted part is
\begin{equation}
  \vec s_\mathrm{out} =
  \frac{1}{2}
  \frac{c_2(t)^\dagger\vec\Sigma c_2(t)}{c_2(t)^\dagger c_2(t)} \,.
\end{equation}
Employing the time evolution operator
\eqref{eq:dirac-second_order_perturbation-theory-result} we find 
\begin{equation}
  \vec s_\mathrm{out} =
  \frac{1}{2}
  \frac{c_0^{+\dagger}(0)U^{++\dagger}_{\mathrm{nd};2, 0}\vec\sigma U^{++}_{\mathrm{nd};2, 0}c_0^+(0)}
  {c_0^{+\dagger}(0)U^{++\dagger}_{\mathrm{nd};2, 0} U^{++}_{\mathrm{nd};2, 0}c_0^+(0)} \,.
\end{equation}
The time evolution operator
\eqref{eq:dirac-second_order_perturbation-theory-result} is up to a
multiplicative factor a unitary $2\times 2$ matrix and, therefore, may
be witten as
\begin{equation}
  \label{eq:SU2-represetation_of_rotations}
  U^{++}_{\mathrm{nd};2, 0} = 
  \sqrt{P}\, \e^{\I \phi} \left[ 
    \cos \left( \frac{\gamma}{2} \right) \mathds{1} - 
    \I \sin \left( \frac{\gamma}{2} \right) \vec n_\mathrm{r} \cdot \vec \sigma 
  \right]
\end{equation}
with the real-valued parameters $\gamma$, $\vec n_\mathrm{r}$, and $P$
which can be determined via equating coefficients in
\eqref{eq:SU2-represetation_of_rotations} and
\eqref{eq:dirac-second_order_perturbation-theory-result}.  The
expression in the square brackets is the SU(2) representation of a
rotation around the rotation axis given by the unit vector $\vec
n_\mathrm{r}$ and the rotation angle $\gamma$.  Thus, the electron's
spin orientation after diffraction $\vec s_\mathrm{out}$ results from a
rotation of $\vec s_\mathrm{in}$ around the axis $\vec n_\mathrm{r}$ by
the angle $\gamma$.

For nonrelativistic electron momenta the parameters of this rotation can
be uniquely identified by equating coefficients in
\eqref{eq:SU2-represetation_of_rotations} and
\eqref{eq:taylored_2_photon_KDE_propagator}.  We find the rotation axis
\begin{equation}
  \vec n_\mathrm{r} = -
  \frac{1}{\sqrt{9 p_E^2 + p_B^2}}
  \begin{pmatrix}
    0 \\ 3 p_E \\ p_B
  \end{pmatrix}
\end{equation}
and the rotation angle
\begin{equation}
  \label{eq:gamma-2-photon_KDE}
  \gamma = 2 \arctan \left(
    \frac{k \sqrt{9 p_E^2 + p_B^2}}{2 m^2 c^2 - p_B^2 - 5 p_E^2} 
  \right)\,.
\end{equation}
Thus, for nonrelativistic momenta $k$, $p_E$, and $p_B$ only very small
spin rotations may be observed in the two-photon Kapitza-Dirac effect.

\section{Three-photon Kapitza-Dirac effect}
\label{sec:three-photon-KDE}

\subsection{Numerical results}

\begin{figure}[b]
  \centering
  \includegraphics[scale=0.45]{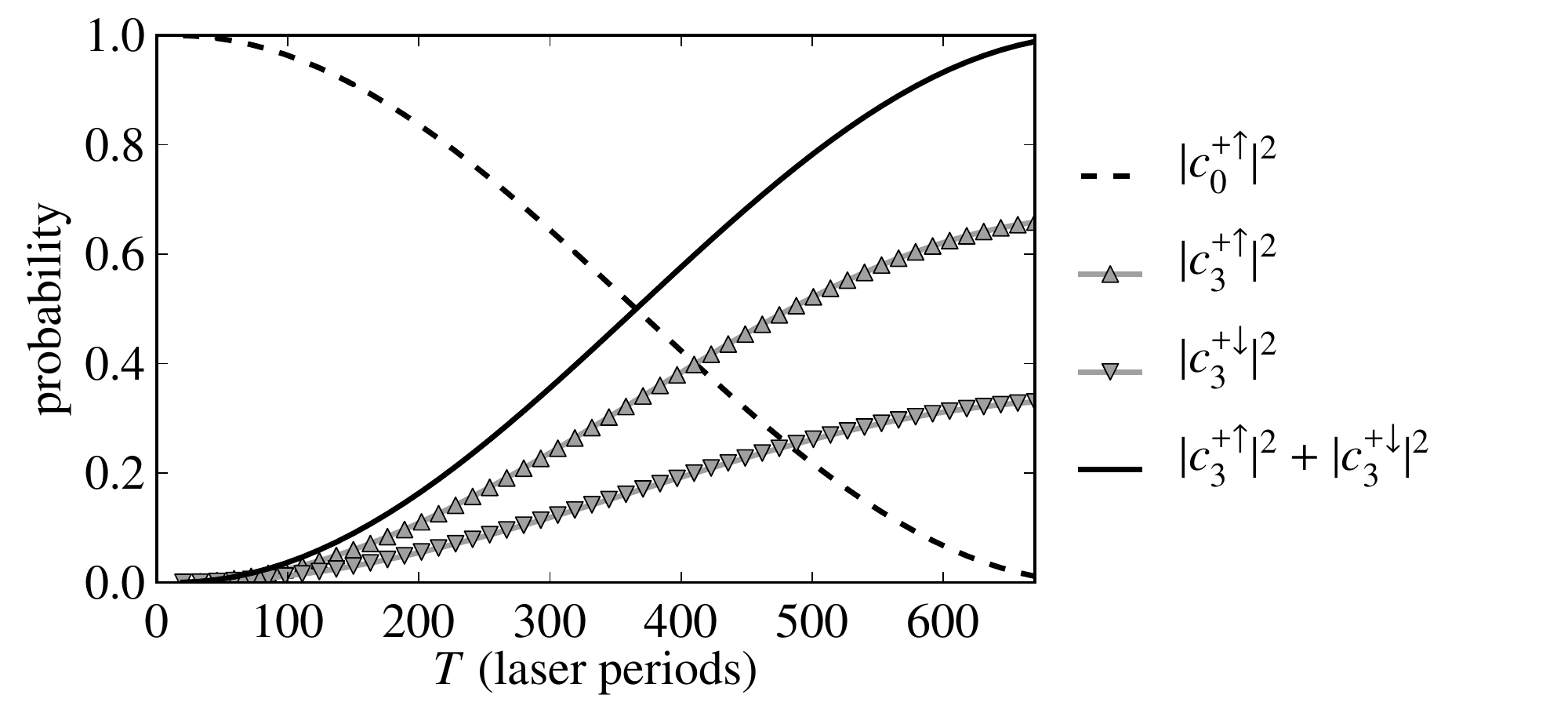}
  \caption{The three-photon Kapitza-Dirac effect simulated by employing
    the Dirac equation \eqref{eq:dirac-equation-momentum-space} for one
    half of a Rabi cycle. The electron with initial momentum
    $176\,\mathrm{keV}=0.347\, m c$ in the laser propagation direction
    (dashed line) is diffracted to the final momentum
    $177\,\mathrm{keV}=0.365\, mc$ in the laser propagation direction
    (solid line) by its interaction with a standing light wave with peak
    intensity $2 \times 10^{23} \mathrm{W}/\mathrm{cm}^2$ (corresponding
    to $e \hat A_\textrm{max} = 0.21 m c^2$) for each beam and photon
    momentum $3.1\,\mathrm{keV}/c=6.1\times 10^{-3}\,mc$. The electron
    momentum in the laser polarization direction is
    $1.2\,\mathrm{keV}=2.4\times 10^{-3}\,mc$. The probability of the
    diffracted electron thereby splits up into a spin-flipped part
    (downward triangles) and a spin-preserving part (upward triangles).
    Data adopted from \cite{Ahrens:etal:2012:spin_kde}.}
  \label{fig:DE_0_3_flip}
\end{figure}

\begin{figure*}
  \centering
  \includegraphics[scale=0.45]{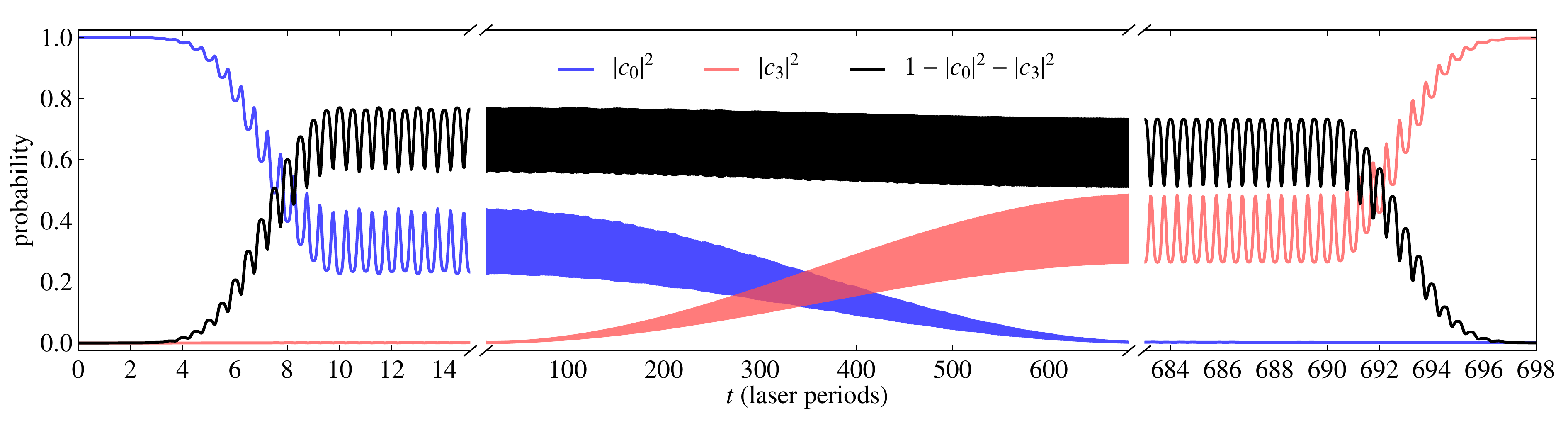}
  \caption{(Color online) The in-field quantum dynamics of the
    three-photon Kapitza-Dirac effect. The graph shows the occupation
    probability of the unscattered beam $|c_0|^2$ and of the diffracted
    beam $|c_3|^2$, and the remaining occupation probability $1 -
    |c_0|^2 - |c_3|^2$ of all other modes. The occupation probabilities
    oscillate in the laser field with twice the laser frequency.  The
    laser field is shaped by a turn-on envelope of ten laser cycles
    (left part of the plot) followed by a period of 678 laser cycles
    with constant laser intensity (center part of the plot) and a
    turn-off envelope of ten laser cycles (right part of the plot); see
    \eqref{eq:turn_on_and_off}.  When the laser is at maximal intensity
    the probability of finding the electron at a certain momentum is
    distributed over the neighboring modes of the modes 0 and 3.  After
    the final turn-off, however, only the modes 0 and 3 are
    occupied. The relative occupation of modes 0 and 3 depends on the
    total interaction time $T$; see Fig.~\ref{fig:DE_0_3_flip}. Here $T$
    has been chosen such that $|c_0(T)|^2\approx0$ and
    $|c_3(T)|^2\approx1$.}
  \label{fig:DE_0_3_flip_dressed}
\end{figure*}

According to the semi-classical considerations in
Sec.~\ref{sec:energy-momentum-conservation}, two photons are absorbed
from the laser field and one photon is emitted into the laser field in
the three-photon Kapitza-Dirac effect, corresponding to $n_r=2$ and
$n_l=-1$.  This particular Kapitza-Dirac effect has been examined in our
recent publication \cite{Ahrens:etal:2012:spin_kde} where we
demonstrated that explicit spin dynamics can be observed depending on
the electron's and the laser's parameters.  Here we will investigate the
three-photon Kapitza-Dirac effect in more detail, analyzing, for
example, the in-field dynamics, conditions for full spin flips, and the
Rabi frequency at relativistic electron momenta.  For the readers'
convenience we show in Fig.~\ref{fig:DE_0_3_flip} the time evolution of
the occupation $|c_n^\zeta(t)|^2$ governed by the Dirac equation in
momentum space \eqref{eq:dirac-equation-momentum-space} for the
parameters that have also been considered in
\cite{Ahrens:etal:2012:spin_kde}.  The parameters of the simulation are
a photon momentum of $k=6.1 \times 10^{-3} m c$, an electron momentum of
$p_E=2.4 \times 10^{-3} m c$ in the laser polarization direction, an
electron momentum of $p_k=0.347\, m c$ in the laser propagation
direction and a field amplitude of $e \hat A_\textrm{max} = 0.21 m c^2$.
Starting from the initial condition \eqref{eq:initial_condition} the
quantum dynamics features Rabi oscillations in the form
\begin{subequations}
  \label{eq:3-photon_KDE-diffraction_probability}
  \begin{align}
    \label{eq:3-photon_KDE_rabioscillation_cos}
    |c_{0}^{+\uparrow}(T)|^2 & = 
    \cos^2 \left(\frac{\Omega_R T}{2} \right) \,,\\
    \label{eq:3-photon_KDE_rabioscillation_sin}
    |c_{3}^{+\uparrow}(T)|^2 + |c_{3}^{+\downarrow}(T)|^2 &=  
    \sin^2 \left(\frac{\Omega_R T}{2} \right)\,,
  \end{align}
\end{subequations}
similarly to the two-photon Kapitza-Dirac effect. However, a fraction of
$0.33$ of the diffracted electrons have flipped their spin with respect
to the quantization axis in the laser polarization direction. 

In contrast to the quantum state after the interaction with the laser
(shown in Fig.~\ref{fig:DE_0_3_flip}), also the neighboring modes of the
modes 0 and 3 are occupied during the in-field quantum dynamics; see
Fig.~\ref{fig:DE_0_3_flip_dressed}.  During the turn-on phase the
quantum state populates the neighboring modes of mode~0 and in the
following interaction with the laser at constant intensity mode~3 gets
partly occupied. After turn-off, however, only the modes~0 and~3 have a
significant occupation probability and all other occupation
probabilities vanish.\vspace*{0pt}

\relax\vspace{1ex plus 1ex}

\subsection{Perturbation theory}
\label{eq:perturbation-theory_3-photon_KDE}

In this section we derive a perturbative short-time solution of the
three-photon Kapitza-Dirac effect, for obtaining analytic expressions
for the Rabi frequency and the electron spin-flip probability, as in the
two-photon Kapitza-Dirac effect. Analogously to the two-photon
Kapitza-Dirac effect, we want to approximate the propagator
$U_{3,0}(t,0)$, which maps the initial quantum state $c_0(0)$ to the
final quantum state by $c_3(t)$ by
\begin{equation}
  \label{eq:c_3_propagation}
  c_3(t) = U_{3,0}(t, 0) c_0(0)\,.
\end{equation}
Since the Dirac equation contains only couplings to the next-neighboring
modes, the lowest-order perturbative solution is of third order and
reads
\begin{widetext}
  \begin{equation}
    U_\mathrm{rd}(t,0) = 
    \frac{1}{\I^3} \int_{0}^{t} \D t_3 \int_{0}^{t_3} \D t_2 \int_{0}^{t_2} \D t_1 U_0(t,t_3) V(t_3) 
    U_0(t_3,t_2)V(t_2) U_0(t_2,t_1) V(t_1) U_0(t_1,0)\,.
    \label{eq:third-order-perturbation-theory}
  \end{equation}
  Utilizing the explicit form of the free propagator
  \eqref{eq:free-propagator} and the interaction Hamiltonian
  \eqref{eq:interaction-hamiltonian} in momentum space yields the
  time-dependent perturbation theory propagator\pagebreak[2]
  \begin{multline}
    U_{\mathrm{rd};3,0}(t,0) = \frac{\I q^3}{8} L_{3,2}^{++} L_{2,1}^{++} L_{1,0}^{++} \int_{0}^{t} \D t_3 \int_{0}^{t_3} \D t_2 \int_{0}^{t_2} \D t_1 \sin(\omega t_3) \sin(\omega t_2) \sin(\omega t_1) \nu^{++}(t,t_3,t_2,t_1) \\
    + \frac{\I q^3}{8} L_{3,2}^{+-} L_{2,1}^{-+} L_{1,0}^{++} \int_{0}^{t} \D t_3 \int_{0}^{t_3} \D t_2 \int_{0}^{t_2} \D t_1 \sin(\omega t_3) \sin(\omega t_2) \sin(\omega t_1) \nu^{-+}(t,t_3,t_2,t_1) \\
    + \frac{\I q^3}{8} L_{3,2}^{++} L_{2,1}^{+-} L_{1,0}^{-+} \int_{0}^{t} \D t_3 \int_{0}^{t_3} \D t_2 \int_{0}^{t_2} \D t_1 \sin(\omega t_3) \sin(\omega t_2) \sin(\omega t_1) \nu^{+-}(t,t_3,t_2,t_1) \\
    + \frac{\I q^3}{8} L_{3,2}^{+-} L_{2,1}^{--} L_{1,0}^{-+} \int_{0}^{t} \D t_3 \int_{0}^{t_3} \D t_2 \int_{0}^{t_2} \D t_1 \sin(\omega t_3) \sin(\omega t_2) \sin(\omega t_1) \nu^{--}(t,t_3,t_2,t_1) \,, \label{eq:dirac_propagator_third_order_perturbation_theory_expanded}
  \end{multline}
\end{widetext}
with the phases
\begin{multline}
  \label{eq:dirac_third_order_perturbation_theory_phase}
  \nu^{ab}(t,t_3,t_2,t_1) = \\
  \exp \left[
    - \I \left( 
      \E_3 t + \Delta \E_{2,3}^{a+} t_3 + \Delta \E_{1,2}^{ba} t_2  +
      \Delta \E_{0,1}^{+b} t_1 
    \right) 
  \right]\,,
\end{multline}
where the upper indices $a$ and $b$ take the values $+$ and
$-$. Analogously to Sec.~\ref{eq:2-photon_KDE_perturbation_theory}, the
integrand of the time integral
\begin{equation}
 \int_{0}^{t} \D t_3 \int_{0}^{t_3} \D t_2 \int_{0}^{t_2} \D t_1 
 \sin(\omega t_3) \sin(\omega t_2) \sin(\omega t_1) \nu^{ab}(t,t_3,t_2,t_1)
\end{equation}
also contains oscillating terms which become constant if the resonance
condition of the three-photon Kapitza-Dirac effect $\E_3 - \E_0 =
\omega$ is met. Accounting only for the terms which grow linearly in
time yields the propagator
\begin{multline}
  \label{eq:dirac-third_order_perturbation-theory-result}
  U_{\mathrm{rd};3,0}^{++}(t,0) = \\
  \left(- \frac{q^3 t}{64}\right) \e^{-\I \E_3 t}
  \left(
    l^{++} L_{3,2}^{++} L_{2,1}^{++} L_{1,0}^{++} + l^{-+}  L_{3,2}^{+-} L_{2,1}^{-+} L_{1,0}^{++} 
  \right.\\
  \left.
    \mbox{}
    + l^{+-} L_{3,2}^{++} L_{2,1}^{+-} L_{1,0}^{-+} + l^{--}  L_{3,2}^{+-} L_{2,1}^{--} L_{1,0}^{-+}
  \right)\,,
\end{multline}
with the coefficients
\begin{multline}
  l^{ab} = 
  \frac{1}{\Delta E_{0,1}^{+b} - \omega}
  \frac{1}{\Delta E_{0,2}^{+a}} + 
  \frac{1}{\Delta E_{0,1}^{+b} + \omega}
  \frac{1}{\Delta E_{0,2}^{+a}}\\
  + \frac{1}{\Delta E_{0,1}^{+b} + \omega}
  \frac{1}{\Delta E_{0,2}^{+a} + 2 \omega} \,.
\end{multline}

\subsection{The relativistic Rabi frequency}

\begin{figure}
  \centering
  \includegraphics[scale=0.45]{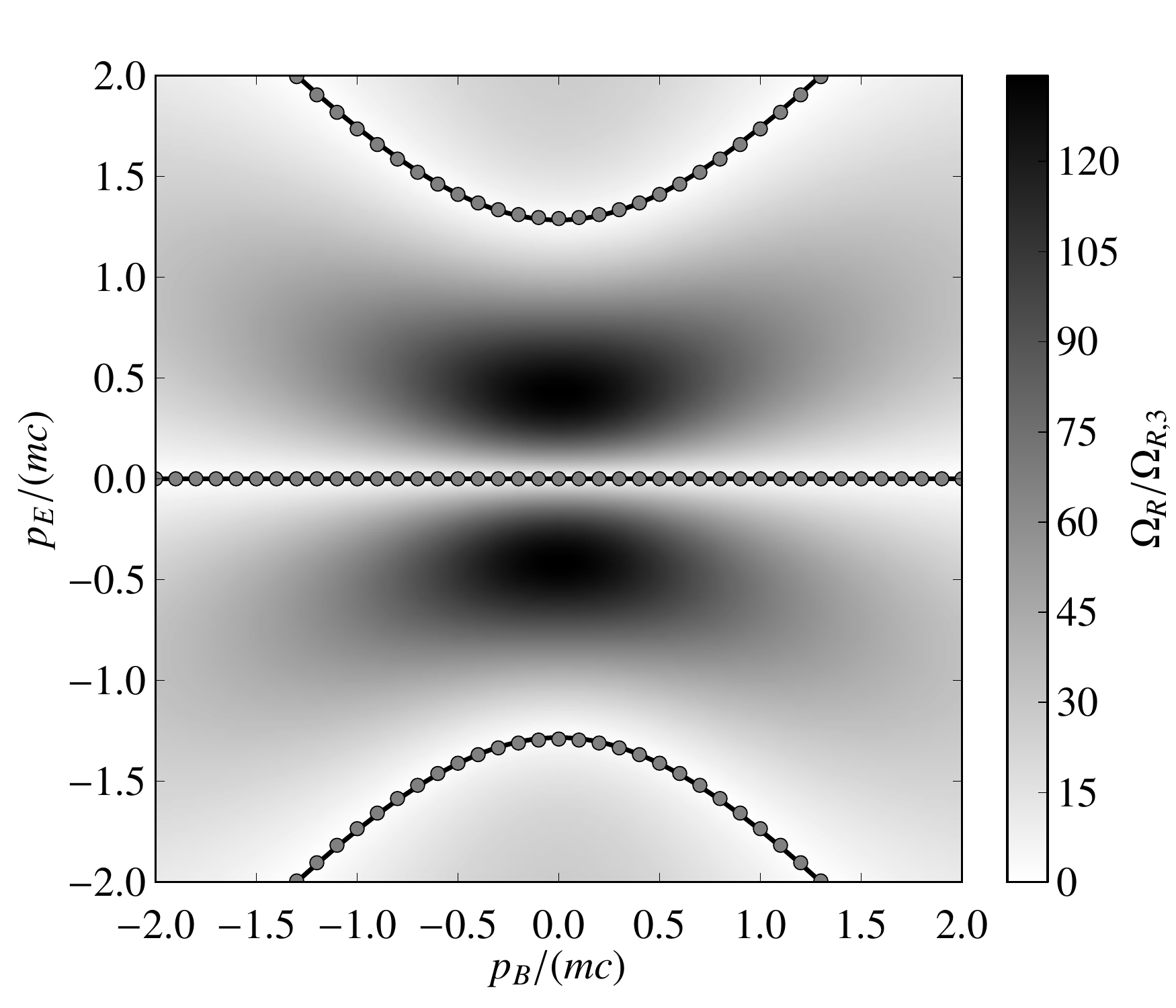}
  \caption{The relativistic Rabi frequency of the three-photon
    Kapitza-Dirac effect normalized to the full spin-flip Rabi frequency
    $\Omega_{R,3}$ in Eq.~\eqref{eq:full-spin-flip-Rabi-frequency}. The
    parameters in this figure are a laser intensity of $2 \times
    10^{23}\,\textrm{W}/\textrm{cm}^2$ and a photon energy of
    3.1\,keV\,. The gray filled circles mark positions in which
    condition \eqref{eq:propagator_3-photon_KDE spin-flip_condition} is
    fulfilled, which are approximated by the
    Eqs.~\eqref{eq:3-photon_spin-flip_condition}.}
  \label{fig:3-photon_KDE-spin-flip_condition}
\end{figure}

An expansion of the propagator $U_{\mathrm{rd};3,0}^{++}(t,0)$ in
Eq.~\eqref{eq:dirac-third_order_perturbation-theory-result} with respect
to the momenta $k$, $p_B$ and $p_E$ yields
\begin{multline}
  \label{eq:dirac-third_order_perturbation-theory_Taylored}
  U_{\mathrm{rd};3,0}^{++}(t,0) = \\
  \frac{q^3 t}{48} \frac{\hat A_\mathrm{max}^3}{m^2 c^4} 
  \frac{\e^{-\I \E_3 t}}{m c}\left( 
    \frac{5 p_E}{\sqrt{2}} \mathds{1} - \I k \sigma_B - 
    \I \frac{ 3 - 2 \sqrt{2}}{m c} k \, p_B \sigma_k 
  \right) \\
  + \mathcal{O}(k^3,p_E^3,p_B^3)\,.
\end{multline}
Evaluating $|c_3^{+\uparrow}(t)|^2 + |c_3^{+\downarrow}(t)|^2$ in
analogy to \eqref{eq:norm_c_2_t} with \eqref{eq:c_3_propagation} and
\eqref{eq:dirac-third_order_perturbation-theory_Taylored} yields the
short-time diffraction probability
\begin{multline}
  |c_3^{+\uparrow}(t)|^2 + |c_3^{+\downarrow}(t)|^2 = \\
  \frac{t^2}{4} \left(\frac{q^3 \hat E^3}{24 m^3 c^5 k^3}\right)^2  
  \left[ \frac{25}{2} p_E^2 + k^2 + \left( 3 - 2 \sqrt{2} \right) k^2 \frac{p_B^2}{m^2 c^2} \right]\,.
\end{multline}
By comparing this probability with the analogous short-time expansion
of the ansatz \eqref{eq:2-photon_KDE_rabioscillation-ansatz}, one finds
the Rabi frequency
\begin{equation}
  \label{eq:Rabi_DE}
  \Omega_R = \Omega_{R,3}
  \sqrt{\frac{25}{2} \frac{p_E^2}{k^2} + 1 + \left( 3 - 2 \sqrt{2} \right) \frac{p_B^2}{m^2 c^2}}\,,
\end{equation}
with the Rabi frequency
\begin{equation}
 \label{eq:full-spin-flip-Rabi-frequency}
 \Omega_{R,3}=\frac{q^3 \hat E^3}{24 m^3 c^5 k^2} \,.
\end{equation}
The expression \eqref{eq:Rabi_DE} holds only for small $k$, $p_E$, and
$p_B$. However, one may evaluate $|c_3^{+\uparrow}(t)|^2 +
|c_3^{+\downarrow}(t)|^2$ with the exact relativistic propagator
\eqref{eq:dirac-third_order_perturbation-theory-result}, which is shown
in Fig.~\ref{fig:3-photon_KDE-spin-flip_condition}. The common property
of the Rabi frequencies $\Omega_{R,2}$ and $\Omega_{R,3}$ is that both
hold at the origin $p_E=p_B=0$ of the $p_E$-$p_B$ plane. The difference
between the two is that the two-photon Kapitza-Dirac effect shows no
spin flip for $p_E=p_B=0$ and the Rabi frequency is maximal for these
momenta. In the three-photon Kapitza-Dirac effect $p_E=p_B=0$ implies a
full spin-flip position (see the next section) and the Rabi frequency
has a saddle point. For the parameters which are used in
Fig.~\ref{fig:DE_0_3_flip} this Rabi frequency evaluates to
$\Omega_R=3.43\times10^{15}\,\textrm{Hz}$ whereas the Rabi frequency
from the simulation $\Omega_R=3.34\times10^{15}\,\textrm{Hz}$ agrees
well with the analytical result.

In analogy to the two-photon Kapitza-Dirac effect, the off-resonant
diffraction probability \eqref{eq:2_off_resonant_scattering_prob} with
the off-resonant Rabi frequency
\eqref{eq:generalized-detuning-rabi-frequency-relation} also applies to
the three-photon Kapitza-Dirac effect. For the parameters applied in
Fig.~\ref{fig:DE_0_3_flip} we find numerically $b=45.7$. We remark that
in the case of the three-photon Kapitza-Dirac effect a systematic shift
of the resonance peak appears in the numerical simulation, as compared
to the peak position which we obtain from the classical resonance
condition \eqref{eq:p_k_relativistic}. For the parameters
$p_k=0.3470\,mc$ and $p_E=2.4 \times 10^{-3}\,mc$ of
Fig.~\ref{fig:DE_0_3_flip} for example, one finds the resonance peak at
the photon momentum $k=6.1 \times 10^{-3}\,mc$ in the numerical
simulation, whereas the resonance condition \eqref{eq:p_k_relativistic}
predicts the photon momentum $k=4.4 \times 10^{-3}\,mc$. This shift
scales with the laser intensity, as the quantum dynamics leaves the
perturbative regime with increasing field amplitude. Therefore, the
resonance peak position of the numerical solution and the classical
condition \eqref{eq:p_k_relativistic} converge to the same value in the
limit of small laser intensities.

\subsection{Spin flips}

\begin{figure}
  \centering
  \includegraphics[scale=0.45]{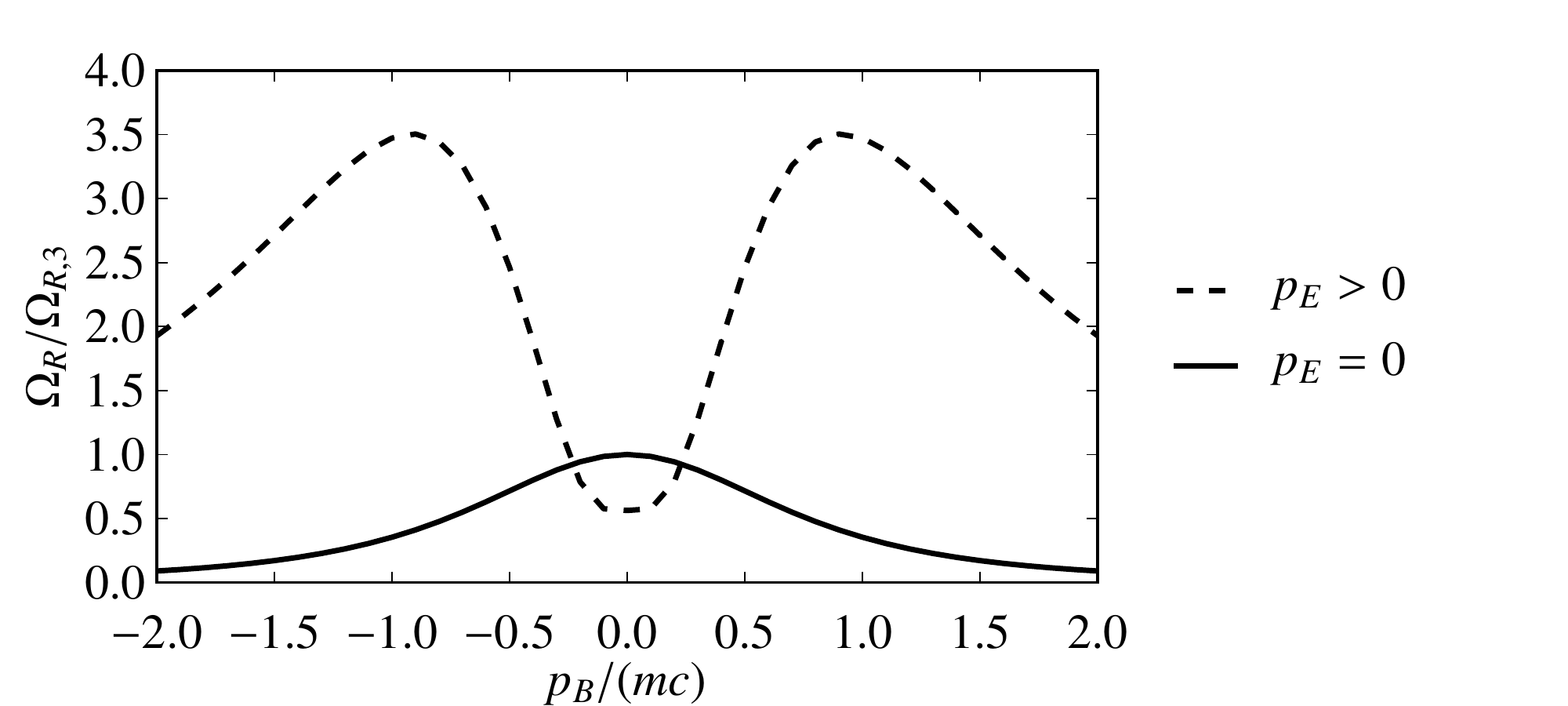}
  \caption{Relativistic Rabi frequency of the three-photon Kapitza-Dirac
    effect at momenta for which full spin-flips are possible (that means
    along the gray filled circles in
    Fig.~\ref{fig:3-photon_KDE-spin-flip_condition}).}
  \label{fig:3-photon-spinflip_Rabi-frequency}
\end{figure}

As in Sec.~\ref{sec:two_photon_spin_flip}, we identify the parameter
space for spin-flips in the three-photon Kapitza-Dirac effect by the
condition
\begin{equation}
  \label{eq:propagator_3-photon_KDE spin-flip_condition}
  \tr\left(\mathds{1}^\dagger U_{\mathrm{rd};3,0}^{++}(t,0)\right)=0\,.
\end{equation}
The numerical evaluation of this condition yields the gray filled
circles in Fig. \ref{fig:3-photon_KDE-spin-flip_condition} which form a
line and a hyperbola.  The line and the hyperbola can be approximated by
the equations
\begin{subequations}
  \label{eq:3-photon_spin-flip_condition}
  \begin{align}
    \label{eq:3-photon_spin-flip_condition_p_E_eq_0}
    p_E &=0\,,\\
    \label{eq:3-photon_spin-flip_condition_p_E_neq_0}
    m^2 c^2 + 0.847 p_B^2 - 0.608 p_E^2 & = 0\,.
  \end{align}
\end{subequations}
The line \eqref{eq:3-photon_spin-flip_condition_p_E_eq_0} can also be
derived from the expansion
\eqref{eq:dirac-third_order_perturbation-theory_Taylored}, because the
spin-flip condition runs through the point $p_E=p_B=0$, and the Taylor
expansion is exact in the vicinity of this point. We also plot the Rabi
frequency at the regions of a full spin flip in
Fig. \ref{fig:3-photon-spinflip_Rabi-frequency}. as in
Sec.~\ref{sec:two_photon_spin_flip}, the Rabi frequency of the spin-flip
regions is about two orders of magnitude lower than the Rabi frequency
of the spin-preserving regions.

\begin{figure}
  \centering
  \includegraphics[scale=0.45]{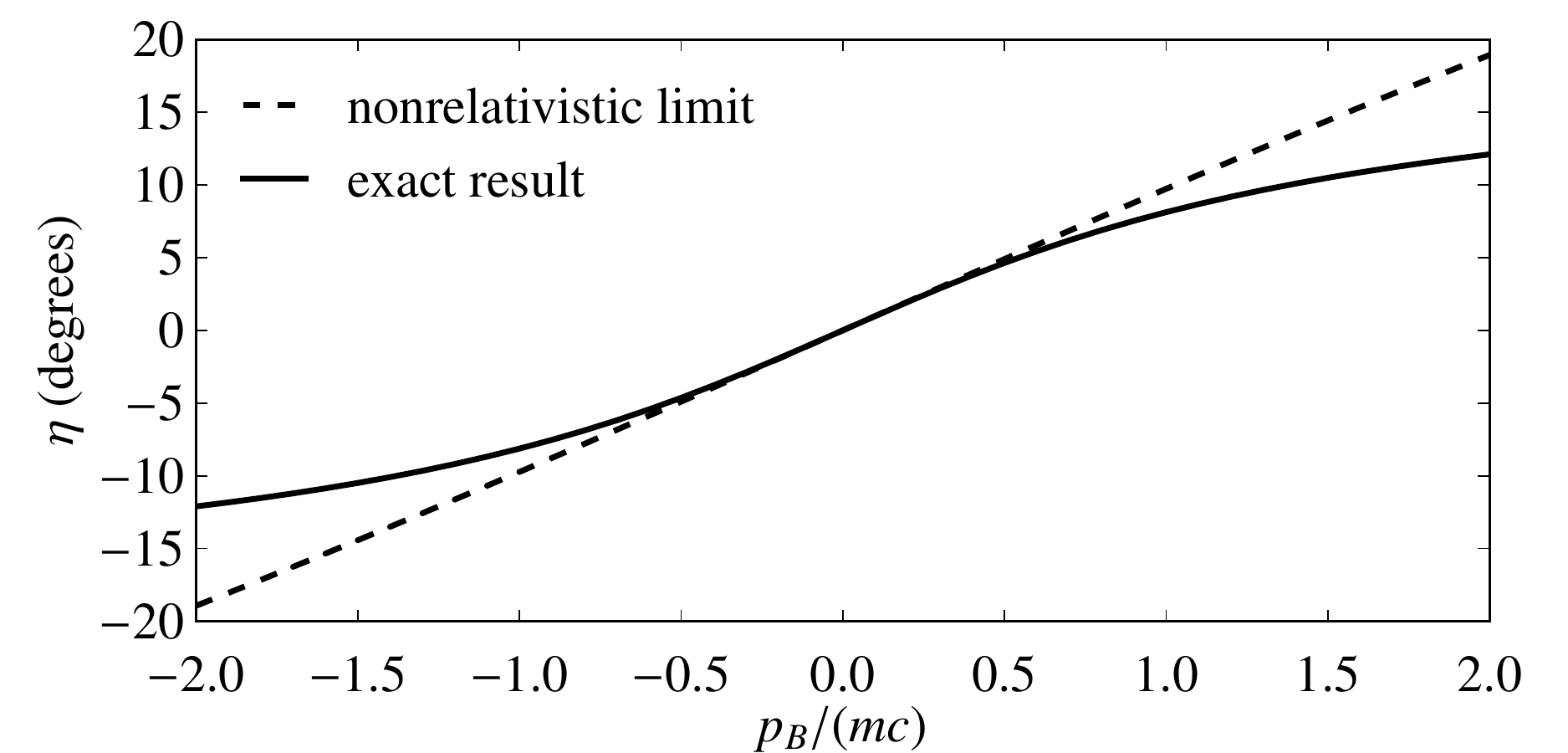}
  \caption{Tilt of the electron spin rotation axis $\vec
    n_\mathrm{r}$. The dashed line shows the analytical result
    \eqref{eq:rotation_axis_tilt} of the expanded propagator
    \eqref{eq:dirac-third_order_perturbation-theory_Taylored}. The black
    line is obtained numerically from the exact propagator
    \eqref{eq:dirac-third_order_perturbation-theory-result}. The tilt of
    the rotation angle takes place in the plane spanned by $k$ and $\hat
    B$, because the $E$ component of the vector
    \eqref{eq:axis-3-photon_KDE} and the $E$ component of the vector
    from the exact computation are zero.}
  \label{fig:3-photon-rotation_axis_angle}
\end{figure}

In analogy to the two-photon Kapitza-Dirac effect the electron spin in the
diffracted part of the laser beam can be expressed as
\begin{equation}
  \vec s_\mathrm{out} =
  \frac{1}{2}
  \frac{c_0^{+\dagger}(0)U^{++\dagger}_{\mathrm{rd};3, 0}\vec\sigma U^{++}_{\mathrm{rd};3, 0}c_0^+(0)}
  {c_0^{+\dagger}(0)U^{++\dagger}_{\mathrm{rd};3, 0} U^{++}_{\mathrm{rd};3, 0}c_0^+(0)} \,.
\end{equation}
by utilizing the time evolution operator
\eqref{eq:dirac-third_order_perturbation-theory-result}.  For
nonrelativistic electron momenta $p_B$ and $p_E$ the parameters of the
rotation of the spin can be uniquely identified by equating coefficients
in \eqref{eq:SU2-represetation_of_rotations} and
\eqref{eq:dirac-third_order_perturbation-theory_Taylored}. In contrast
to the two-photon Kapitza-Dirac effect we find significant spin
rotations for the electron momenta $p_E$ and $p_B$. For the case of the
three-photon Kapitza-Dirac effect, the rotation axis is given by
\begin{equation}
  \label{eq:axis-3-photon_KDE}
  \vec n_\mathrm{r} =
  \frac{1}{\sqrt{\left(3 - 2 \sqrt{2}\right)^2 p_B^2 + m^2 c^2}}
  \begin{pmatrix}
    \left(3  - 2 \sqrt{2}\right) p_B \\ m c \\ 0
  \end{pmatrix}
\end{equation}
and the rotation angle reads
\begin{equation}
  \label{eq:gamma-3-photon_KDE}
  \gamma = 2 \arctan \left[ 
    \frac{\sqrt{2}}{5}\frac{k}{p_E} 
    \sqrt{1 + \left(3 - 2 \sqrt{2}\right)^2 \frac{p_B^2}{m^2 c^2}}
  \right]\,.
\end{equation}
Equation \eqref{eq:axis-3-photon_KDE} means that the electron spin is
rotated around the axis of the magnetic field, if the electron momentum
$p_B$ in the magnetic field direction is zero. A nonvanishing $p_B$,
however, tilts the rotation axis of the spin rotation into the direction
of the laser propagation direction by the angle
\begin{equation}
  \label{eq:rotation_axis_tilt}
 \eta \equiv \arctan\left(
   \frac{n_{\mathrm{r},k}}{n_{\mathrm{r},B}}
 \right) = 
 \arctan \left[ \left(3 - 2 \sqrt{2}\right) \frac{p_B}{m c} \right]\,.
\end{equation}
This tilt angle $\eta$ is plotted together with the exact value,
obtained from the propagator
\eqref{eq:dirac-third_order_perturbation-theory-result}, in
Fig.~\ref{fig:3-photon-rotation_axis_angle}.

\section{Conclusions and Outlook}

We discussed the general relativistic and nonrelativistic Bragg
conditions for a Kapitza-Dirac effect with $n$ interacting photons that
follow from energy-momentum conservation. We investigated the two-photon
and three-photon Kapitza-Dirac effects by accounting for the Bragg
conditions and simulating the Pauli equation and the Dirac equation,
which were transformed into a momentum-space representation. In both
scenarios spin effects appear at relativistic momenta of the
electron. We also presented an analytic solution of the Kapitza-Dirac
effect by computing the short-time evolution with time-dependent
perturbation theory for the two-photon and the three-photon
Kapitza-Dirac effects. By the help of our analytical and numerical
methods, we were able to demonstrate full spin flips within the
two-photon Kapitza-Dirac effect. Furthermore, we pointed out that the
spin flip in the Kapitza-Dirac effect corresponds to a rotation of the
electron spin, when the electron is diffracted.  The diffraction
probability, however, does not depend on the spin orientation of the
incident electrons.  Our numerical simulations indicate that the
off-resonant Kapitza-Dirac effect can be described by a generalization
of Rabi theory for two-level systems.

The Rabi frequency of the $n$-photon Kapitza-Dirac effect scales with
the $n${th} power of $e \hat E/(k m c^2)$, because the lowest order
contribution in time-dependent perturbation theory is of $n${th}
order and contains a product of $n$ times the interaction Hamiltonian
\eqref{eq:interaction-hamiltonian}. Therefore, the Rabi frequency of the
three-photon Kapitza-Dirac effect is suppressed by a factor of $e \hat
E/(k m c^2)$ as compared to the Rabi frequency of the two-photon
Kapitza-Dirac effect. Note that $e \hat E/k$ is always smaller than $m
c^2$ in the Bragg regime. Thus, from this point of view, higher laser
intensities are in principle required for higher photon processes.

The two-photon and three-photon Kapitza-Dirac effects are different, if
the two are compared against the background of spontaneous
emission. Since spontaneous emission is proportional to the square of
the electric field, the radiation power spontaneously emitted by the
electron scales like the Rabi frequency of the two-photon Kapitza-Dirac
effect with the square of $e \hat E/(km c^2)$. This implies that the
spontaneously emitted energy which is emitted in one Rabi cycle is
independent of the laser intensity for the two-photon Kapitza-Dirac
effect. The Rabi frequency of the three-photon Kapitza-Dirac effect,
however, scales with the third power of $e \hat E/(km c^2)$. Therefore,
the three-photon Kapitza-Dirac effect and all higher-order Kapitza-Dirac
effects may become only visible for very high intensities of the
external laser field.

Even though the above considerations favor the two-photon Kapitza-Dirac
effect for an experimental demonstration of spin effects, the two-photon
Kapitza-Dirac effect has the drawback, that spin effects occur only for
relativistic momenta of the injected electron in the laser polarization
direction. This implies less favorable short interaction times of the
electron with the laser, if the electron passes through a narrowly
focused laser beam.

Our numerical solution of the quantum dynamics also shows that many
modes are excited in the in-field dynamics, indicating that the results
from perturbation theory might be applicable even for interaction
parameters in which higher-order perturbative corrections should be of
relevance. The good agreement of our numerical results with the
perturbative approximation moreover suggests the applicability of our
predictions in the parameter space of intense, optical laser
beams. Therefore we conjecture that spin signatures in the Kapitza-Dirac
effect might be realizable even for the interaction of moderately
relativistic electrons with intense laser beams in the optical regime.


\begin{acknowledgments}
  S.~A. would like to thank Rainer Grobe for inspiring discussions and
  Matthias Dellweg for carefully reading the manuscript.
\end{acknowledgments}

\appendix

\section{Spin-interaction matrices}
\label{sec:appendix_spin-matrices}

To express the spin-interaction matrices $L_{n,n'}^{ab}$
\eqref{eq:spin_interaction_matices}, we first introduce the coefficients
\begin{subequations}
  \begin{align}
    d_n^+ & = 
    \frac{1}{\sqrt{2}}\sqrt{\frac{\E_n + mc^2}{\E_n}}\\
    d_n^- & = 
    \frac{c}{\sqrt{2}}\sqrt{\frac{1}{\E_n (\E_n + m c^2)}}\,,
  \end{align}
\end{subequations}
which allows us to define the coefficients
\begin{subequations}
  \label{eq:dirac-spinor-coefficients}
  \begin{align}
    t_{n,n'} & = 
    d^+_n d^+_{n'} + \vec p_n \cdot \vec p_{n'}\, d^-_n d^-_{n'}\,, 
    \label{eq:dirac-spinor-coefficient-t}\\
    s^l_{n,n'}& = 
    p_n^{l} d^-_n d^+_{n'} + p_m^{l} d^+_n d^-_{n'} \,,
    \label{eq:dirac-spinor-coefficient-s}\\
    r^l_{n,n'}& =
    p_n^{l} d^-_n d^+_{n'} - p_{n'}^{l} d^+_n d^-_{n'} \,,
    \label{eq:dirac-spinor-coefficient-r}\\
    w^{l,q}_{n,n'} & =
    p_n^{l} p_{n'}^{q} d^-_n d^-_{n'} + p_n^{q} p_{n'}^{l} d^-_n d^-_{n'}\,,
    \label{eq:dirac-spinor-coefficient-w}\\
    h^{l}_{n,n'} & = 
    \vec e_l \cdot \left(\vec p_n \times \vec p_{n'}\right) d^-_n d^-_{n'}\,.
    \label{eq:dirac-spinor-coefficient-h}
  \end{align}
\end{subequations}
The upper indices denote the vector components of the coefficients,
whereas the lower indices correspond to a mode number of the
wave function \eqref{eq:pauli-wave-function} of the electron. One can
show \cite{Ahrens:2012:PhD_thesis} that the matrices $L_{n,n'}^{ab}$ in
\eqref{eq:spin_interaction_matices} are given by
\begin{multline}
  \label{eq:dirac-energy-keep-matrix}
  L_{n,n'}^{++} = - L_{n,n'}^{--} 
  = \sum_l \hat A^l(t) s^l_{n,n'} \,  \mathds{1} + 
  \I \sum_{lqj} \epsilon_{jlq} r^l_{n,n'} \hat A^q(t) \sigma^j 
\end{multline}
and
\begin{multline}
  \label{eq:dirac-energy-change-matrix}
  L_{n,n'}^{+-} = \phantom{-} L_{n,n'}^{-+} \\
  = \sum_l t_{n,n'} \hat A^l(t) \sigma^l - 
  \sum_{l,q} w^{l,q}_{n,n'} \hat A^q(t) \,\sigma^l + 
  \I \sum_l \hat A^l(t) h^l_{n,n'} \,\mathds{1}\,. 
\end{multline}

\bibliography{detailed_kde}

\end{document}